\documentclass[12pt]{article}%
\usepackage[left=2.5cm, right=2.5cm, top=2.25cm, bottom=2.5cm]{geometry}
\usepackage{natbib}
\usepackage{amsfonts}
\usepackage{amssymb}
\usepackage{amsmath}
\usepackage{lscape}
\usepackage{color}
\usepackage{algorithm}
\usepackage[noend]{algpseudocode}
\usepackage{graphicx}
\usepackage{pgfplots}
\usepackage{pgfplotstable}
\usepackage{epstopdf}
\usepackage{tikz}%
\providecommand{\U}[1]{\protect\rule{.1in}{.1in}}
\pgfplotsset{compat=1.8}
\usepgfplotslibrary{groupplots}
\pgfplotsset{compat=newest}
\pgfplotsset{plot coordinates/math parser=false}
\newlength\figureheight
\newlength\figurewidth
\pgfplotsset{yticklabel style={text width=3em,align=right}}
\pgfplotsset{xticklabel style={text width=2em,align=right}}
\newtheorem{theorem}{Theorem}

\newtheorem{assumption}{Assumption}
\newenvironment{proof}[1][Proof]{\noindent \textbf{#1.} }{\  \rule{0.5em}{0.5em}}

\newcommand{\y}{\mathbf{y}}
\newcommand{\z}{\mathbf{z}}
\makeatletter
\newcommand{\blind}{1}
\newcommand{\T}{\mathbf{\theta}}

\def\1{1\!{\rm l}}
\def\BState{\State\hskip-\ALG@thistlm}
\makeatother
\date{\vspace{-5ex}}
\begin{document}

\if1\blind
{
  \title{Approximate Bayesian Forecasting\thanks{We thank the Editor and two anonymous referees for very constructive and detailed comments on an earlier draft of the paper. This research has been supported by
  Australian Research Council Discovery Grants No. {DP150101728 and
  DP170100729}. }}
  \author{David T. Frazier\thanks{Department of Econometrics and Business Statistics,
  Monash University, Australia. Corresponding author; email: david.frazier@monash.edu.}, Worapree Maneesoonthorn\thanks{Melbourne
  Business School, University of Melbourne, Australia.}, Gael M.
  Martin\thanks{Department of Econometrics and Business Statistics, Monash
  University, Australia.}
  \and and Brendan P.M. McCabe\thanks{Management School, University of Liverpool,
  U.K.}}
  
  \maketitle
} \fi

\if0\blind
{
\title{\LARGE\bf Approximate Bayesian Forecasting\thanks{We thank the Editor and two anonymous referees for very constructive and detailed comments on an earlier draft of the paper.}}
\maketitle

} \fi

\begin{abstract}
Approximate Bayesian Computation (ABC) has become increasingly prominent as a
method for conducting {parameter} inference in a range of challenging
statistical problems, most notably those characterized by an intractable
likelihood function. In this paper, we focus on the use of ABC not as a tool
for parametric inference, but as a means of generating probabilistic
forecasts; or for conducting what we refer to as `approximate Bayesian
forecasting'. The four key issues explored are: {i) the link between the
theoretical behavior of the ABC posterior and that of the ABC-based
predictive; }ii) the {use of proper scoring rules to measure the }%
({potential}) loss of forecast accuracy {when using an approximate rather than
an exact predictive}; iii) the {performance of } approximate Bayesian
forecasting in {state space} models; and iv) the use of forecasting criteria
to inform the selection of ABC summaries in empirical settings. The primary
finding of the paper is that ABC can provide a computationally efficient means
of generating probabilistic forecasts that are nearly identical to those
produced by the exact {predictive}, and in a fraction of the time required to
produce {predictions via an }exact {method}.

\bigskip

\end{abstract}


\smallskip

\noindent\emph{MSC2010 Subject Classification}: 62E17, 62F15, 62F12 \smallskip


\newpage

\section{Introduction}

Approximate Bayesian Computation (ABC) has become an increasingly prominent
inferential tool in challenging problems, most notably those characterized by
an intractable likelihood function. ABC requires only that one can\ simulate
pseudo-data from the assumed model, for given draws of the parameters from the
prior. Parameter draws that produce a\ `match' between the pseudo and observed
data - according to a given set of summary statistics, a chosen\textbf{
}metric and a pre-specified tolerance - are retained and used to estimate the
posterior distribution, with the resultant estimate of the exact (but
inaccessible) posterior being conditioned on the summaries used in the
matching. Various guiding principles have been established to select summary
statistics in ABC (see, for instance, \citealp{JM2008},
\citealp{drovandi2015}, and \citealp{FP2012}) and we refer the reader to
reviews by \cite{Blum2013} and \cite{prangle2015summary} for discussions of
these different approaches.

Along with the growth in applications of ABC (see \citealp{marinea2012},
{\citealp{sisson2011likelihood}}, and \citealp{Robert2016}, for recent
surveys), attention has recently been paid to the theoretical {properties} of
the method, {including the asymptotic behaviour of: ABC posterior
distributions, point estimates derived from those distributions, and Bayes
factors that condition on summaries.} Notable contributions here are
\cite{Marinetal2014}, \cite{Creelal2015}, \cite{Jasra2015}, \cite{Martin2017},
\cite{LF2016b}, \cite{LF2016a} and \cite{FMRR2016}, with \cite{FMRR2016}
providing the full suite of asymptotic results pertaining to the ABC posterior
- namely,\textbf{\ }Bayesian (or posterior) consistency, limiting posterior
shape, and the asymptotic distribution of the posterior mean.

This current paper stands in contrast to {the vast majority of} ABC studies,
with their focus on parametric inference and/or model choice. Our goal herein
is to exploit ABC {as a means of }generating probabilistic \textit{forecasts};
or for conducting what we refer to hereafter as `approximate Bayesian
forecasting' (ABF). Whilst ABF has particular relevance in scenarios in which
the likelihood function and, hence, the exact predictive\textbf{\ }%
distribution, is inaccessible, we also give attention to cases
where\textbf{\ }the exact predictive \textit{is} able to be estimated (via {a
Monte Carlo Markov chain algorithm}), {but at a }greater computational cost
than that associated with ABF. That is, in part, we explore ABF as a
computationally convenient means of constructing predictive
distributions.\footnote{Throughout the paper, we use the terms `forecast' and
`prediction', and their various adjectival forms and associated verb
conjugations, synonymously, interchanging them for linguistic variety only.}

We prove that, under certain regularity conditions, ABF produces forecasts
that are asymptotically equivalent to those obtained from exact Bayesian
methods, and {illustrate numerically the close match that can occur between
approximate and exact predictives, even when the corresponding approximate and
exact posteriors }for the parameters {are very distinct. }We also explore the
application of ABF to state space models, in which the production of an
approximate Bayesian predictive requires integration over both a small number
of static parameters and a set of states with dimension equal to the sample size.

{In summary, the four} primary questions addressed {in the paper }are the
following: i) What role does the asymptotic behavior of the ABC posterior - in
particular Bayesian consistency - play in determining the accuracy of the
approximate predictive as an estimate of the exact predictive? ii) Can we
characterize the loss incurred by using the approximate rather than the exact
predictive, using proper scoring rules? iii) How does ABF perform in {state
space} models,\textbf{ }and {what role does }(particle) filtering {play
therein}? iv) How can forecast accuracy be used to guide the choice of summary
statistics in an empirical setting?

We note that independent of this research, \cite{canale2016} propose
the use of ABC as a means of generating nonparametric forecasts of certain
functional time series models\textbf{ }with intractable likelihoods. In
particular, \cite{canale2016} use ABC sampling as a means of generating
$h$-step ahead point and interval\textbf{ }forecasts for some underlying
unknown curve of interest. The authors apply this methodology to the
prediction of price dynamics in the Italian natural gas market. Whilst not
pursuing the same lines of enquiry as in the current research, the \cite{canale2016} paper highlights the usefulness of ABC as a forecasting tool in
scenarios when exact Bayesian inference - and, hence, exact Bayesian
prediction - is infeasible, and thereby provides further evidence of the
practical importance of the results we provide herein.

The remainder of the paper proceeds as follows. In Section 2 we {first
}provide a brief overview of the method of ABC for producing estimates of an
exact, but potentially inaccessible, posterior for the unknown parameters. The
use of an ABC posterior to yield an approximate forecast distribution is then
proposed. {After a brief outline of existing asymptotic results pertaining to
ABC in Section \ref{asym_abc},} the role played by Bayesian consistency in
determining the accuracy of ABF is formally established in Section \ref{merg},
{with this building} on earlier insights by \cite{blackwell1962} and
\cite{diaconis1986} regarding the merging of predictive distributions. In
Section \ref{proper}, the concept of a proper scoring rule is adopted in order
to formalize the loss incurred when adopting the approximate rather than the
exact Bayesian predictive. The relative performance of ABF is then quantified
in Section \ref{numerics 1} using two simple examples: one in which an integer
autoregressive model for count time series data is adopted as the data
generating process (DGP), with a {single set of }summaries used to implement
ABC; and a second in which a moving average (MA) model is the assumed DGP, and
predictives based on alternative sets of summaries are investigated. In both
examples there is little visual distinction between the approximate and exact
predictives, despite enormous visual differences between the corresponding
posteriors. Furthermore, the visual similarity between the exact and
approximate predictives extends to forecast accuracy: using averages of
various proper scores over a hold-out sample, we demonstrate that the
predictive superiority of the exact predictive, over the approximate, is
minimal in both examples. Moreover, we highlight the fact that all approximate
predictives can be produced in a fraction of the time taken to produce the
corresponding exact predictive.

In Section 4, we explore ABF in the context of a model in which latent
variables feature. Using a simple stochastic volatility model for which the
exact predictive is accessible via Markov chain Monte Carlo (MCMC), the
critical importance {(in terms of matching the exact predictive) of augmenting
ABC inference on the static parameters with `exact' inference on the states,
via a particle filtering step, is made clear}. An extensive empirical
illustration is then undertaken in Section 5. Approximate predictives for both
a financial return and its volatility, in a dynamic jump diffusion model with
$\alpha$-stable volatility transitions, are produced, {using different sets of
summary statistics}, including those extracted from simple auxiliary models
with closed-form likelihood functions. Particular focus is given to using
out-of-sample predictive performance to {choose} the `best' set of summaries
for driving ABC, in the case where prediction is the primary goal of the
investigation. A discussion section concludes the paper in Section 6, and
proofs are included in the Appendix. {All Matlab code} used in the production
of the numerical results will be made available at
http://users.monash.edu.au/\symbol{126}gmartin/.

\section{Approximate Bayesian Computation (ABC): Inference and Forecasting}

We observe a $T$-dimensional vector of data $\mathbf{y}=(y_{1},y_{2}%
,...,y_{T})^{\prime}$, assumed to be generated from some model with likelihood
$p(\mathbf{y}|\mathbf{\theta})$, with ${\theta\in\Theta\subseteq}$
$\mathbb{R}^{k_{\theta}}$ a $k_{\theta}$-dimension vector of unknown
parameters, and where we possess prior beliefs on $\mathbf{\theta}$ specified
by $p(\mathbf{\theta})$. In this section, we propose a means of producing
probabilistic forecasts for the random variables $Y_{T+k},k=1,...,h$, in
situations where $p(\mathbf{y}|\mathbf{\theta})$ is computationally
intractable or numerically difficult to calculate. Before presenting this
approach, we first give a brief overview of ABC-based inference for the
unknown parameters $\mathbf{\theta}$.

\subsection{ABC Inference: Overview}

The aim of ABC is to produce draws from an approximation to the posterior
distribution,
\begin{equation}
p(\mathbf{\theta|y})\propto p(\mathbf{y|\theta})p(\mathbf{\theta}),
\label{exact_posterior}%
\end{equation}
in the setting where both $p(\mathbf{\theta})$, and the assumed data
generating process, $p(\mathbf{y|\theta})$, can be simulated from, but
where\textbf{\ }$p(\mathbf{y|\theta})$ is intractable in some sense. These
draws are, in turn, used to approximate posterior quantities of interest, and
thereby form the basis for conducting inference about $\mathbf{\theta}$. The
simplest (accept/reject) form of the algorithm proceeds as in Algorithm
1.\begin{algorithm}
\caption{ABC accept/reject algorithm}\label{ABC}
\begin{algorithmic}[1]
\State Simulate $\mathbf{\theta }^{i}$, $i=1,2,...,N$, from $p(\mathbf{\theta })$
\State Simulate $\mathbf{z}^{i}=(z_{1}^{i},z_{2}^{i},...,z_{T}^{i})^{\prime }$, $i=1,2,...,N$, from the likelihood, $p(\mathbf{.|\theta }^{i})$
\State Select $\mathbf{\theta }^{i}$ such that:%
\begin{equation}
d\{\mathbf{\eta }(\mathbf{y}),\mathbf{\eta }(\mathbf{z}^{i})\}\leq
\varepsilon ,  \label{distance}
\end{equation}%
where $\mathbf{\eta} (\mathbf{.})$ is a (vector) statistic, $d\{.\}$ is a distance
criterion, and, given $N$, the tolerance level $\varepsilon $ is chosen to be small. (The Euclidean distance is used for all numerical illustrations in the paper.)
\end{algorithmic}
\end{algorithm}

The algorithm thus samples $\mathbf{\theta}$ and pseudo-data $\mathbf{z}$ from
the joint posterior:%
\[
p_{\varepsilon}(\mathbf{\theta},\mathbf{z|\eta(y)})=\frac{p(\mathbf{\theta
})p(\mathbf{z|\theta})\mathbb{I}_{\varepsilon}[\mathbf{z}]}{\textstyle\int%
_{\mathbf{\Theta}}\int_{\mathbf{z}}p(\mathbf{\theta})p(\mathbf{z|\theta
})\mathbb{I}_{\varepsilon}[\mathbf{z}]d\mathbf{z}d\mathbf{\theta}},
\]
where $\mathbb{I}_{\varepsilon}[\mathbf{z}]$:=$\mathbb{I}[d\{\mathbf{\eta
}(\mathbf{y}),\mathbf{\eta}(\mathbf{z})\}\leq\varepsilon]$ is one if
$d\left\{  \mathbf{\eta}(\mathbf{y}),\mathbf{\eta}(\mathbf{z})\right\}
\leq\varepsilon$ and zero otherwise. When the vector of summary statistics,
$\mathbf{\eta}(\mathbf{\cdot})$, is sufficient for $\mathbf{\theta}$ and
$\varepsilon$ {is small},
\begin{equation}
p_{\varepsilon}(\mathbf{\theta|\eta(y)})=\textstyle\int_{\mathbf{z}%
}p_{\varepsilon}(\mathbf{\theta},\mathbf{z|\eta(y)})d\mathbf{z}%
\label{abc_post}%
\end{equation}
approximates $p(\mathbf{\theta|y})$ {well,} and draws from $p_{\varepsilon
}(\mathbf{\theta|\eta(y)})$ can be used to estimate features of that exact
posterior. In practice however, the complexity of the models to which ABC is
applied implies that a low-dimensional vector of sufficient statistics does
not exist. Hence, as $\varepsilon\rightarrow0$ the draws can be used to
estimate features of $p(\mathbf{\theta|\eta}(\mathbf{y}))$ only, with the
`proximity' of $p(\mathbf{\theta|\eta}(\mathbf{y}))$ to $p(\mathbf{\theta|y})$
depending - in a sense that is not formally defined - on the `proximity' to
sufficiency of $\mathbf{\eta}(\mathbf{y}).$

{Unlike }most{\textbf{\ }existing studies on ABC,\ our end goal is
\textit{not} the quantification of uncertainty about $\mathbf{\theta}$, but
the construction of probabilistic forecasts for future realizations of a
random variable of interest, }in which{\textbf{\ }}$p_{\varepsilon
}(\mathbf{\theta|\eta}(\mathbf{y}))$ expresses our uncertainty
about\textbf{\ }$\mathbf{\theta.}$ That is, in contrast to exact Bayesian
forecasting, in which a (marginal) predictive distribution is produced by
averaging the conditional predictive with respect to the exact
posterior,{\textbf{\ }}$p(\mathbf{\theta|y})$, approximate Bayesian
forecasting performs this integration step using the approximate posterior as
the weighting function. This substitution (of\textbf{\ }$p(\mathbf{\theta|y})$
by\textbf{\ }$p_{\varepsilon}(\mathbf{\theta|\eta}(\mathbf{y}))$) is most
clearly motivated in cases where\textbf{\ }$p(\mathbf{\theta|y})${\textbf{\ }%
}is inaccessible, due to an intractable likelihood function. However, the use
of{\textbf{\ }}$p_{\varepsilon}(\mathbf{\theta|\eta}(\mathbf{y}))$ will also
be motivated here by computational considerations alone.

\subsection{Approximate Bayesian Forecasting (ABF)}

Without loss of generality, we focus at this point on one-step-ahead
forecasting in the context of a time series model.\footnote{Multi-step-ahead
forecasting entails no additional conceptual challenges and, hence, is not
treated herein.} Let $Y_{T+1}$ denote a random variable that will be observed
at time $T+1$, and which is generated from the (conditional) predictive
density (or mass) function, $p(y_{T+1}|\mathbf{\theta},\mathbf{y})$, at some
fixed value $\mathbf{\theta}$. The quantity of interest is thus%
\begin{equation}
p(y_{T+1}|\mathbf{y})=\int_{\Theta}p(y_{T+1}|\mathbf{\theta},\mathbf{y}%
)p(\mathbf{\theta}|\mathbf{y})d\mathbf{\theta,}\label{exact_predictive}%
\end{equation}
where $p(\mathbf{\theta}|\mathbf{y})$ is the exact posterior defined in
(\ref{exact_posterior}) and $y_{T+1}$ denotes a value in the support of
$Y_{T+1}$. The DGP, $p(\mathbf{y|\theta})$, is required in closed form for
numerical methods such as MCMC to be applicable to $p(\mathbf{\theta
}|\mathbf{y})$, in the typical case in which the latter itself cannot be
expressed in a standard form.\footnote{Pseudo-marginal MCMC methods may be
feasible when certain components of the DGP are unavailable in closed form.
For example, particle MCMC could be applied to state space models in which the
state transitions are unavailable, but can be simulated from. However, the
great majority of MCMC algorithms would {appear to }exploit full knowledge of
the DGP in their construction.} Such methods yield draws from
$p(\mathbf{\theta}|\mathbf{y})$ that are then used to produce a
simulation-based estimate of the predictive density as:%
\begin{equation}
\widehat{p}(y_{T+1}|\mathbf{y})=\frac{1}{M}\sum_{i=1}^{M}p(y_{T+1}%
|\mathbf{\theta}^{(i)},\mathbf{y}),\label{est_pred}%
\end{equation}
where the conditional predictive, $p(y_{T+1}|\mathbf{\theta}^{(i)}%
,\mathbf{y})$, is also required to be known in closed-form for the
`Rao-Blackwellized' estimate in (\ref{est_pred}) to be feasible.
Alternatively, draws of $y_{T+1}$ from $p(y_{T+1}|\mathbf{\theta}%
^{(i)},\mathbf{y})$ can be used to produce a kernel density estimate of
$p(y_{T+1}|\mathbf{y}).$ Subject to convergence of the MCMC chain, either
computation represents an estimate of the exact predictive that is accurate up
to simulation error, and may be referred to as yielding the \textit{exact}
Bayesian forecast distribution as a consequence.

The motivation for the use of ABC in this setting is obvious: in cases where
$p(\mathbf{y|\theta})$ is not accessible, $p(\mathbf{\theta}|\mathbf{y})$
itself is inaccessible (via an MCMC scheme of some sort, {for example}) and
the integral in (\ref{exact_predictive}) that\textbf{\ }defines the exact
predictive cannot be estimated via those MCMC draws in the manner described
above. ABC enables approximate Bayesian \textit{inference} about
$\mathbf{\theta}$ to proceed via a simulation-based estimate of
$p(\mathbf{\theta}|\mathbf{\eta}(\mathbf{y}))$, for some chosen summary,
$\mathbf{\eta}(\mathbf{y}).$ Hence, a natural way in which to approach the
concept of approximate Bayesian \textit{forecasting }is to define the quantity%
\begin{equation}
g(y_{T+1}|\mathbf{y)}=\int_{\Theta}p(y_{T+1}|\mathbf{\theta},\mathbf{y}%
)p_{\varepsilon}(\mathbf{\theta|\eta}(\mathbf{y}))d\mathbf{\theta,}
\label{abc_predictive}%
\end{equation}
with $p_{\varepsilon}(\mathbf{\theta|\eta}(\mathbf{y}))$ replacing
$p(\mathbf{\theta}|\mathbf{y})$ in (\ref{exact_predictive}). The
conditional\textbf{ }density function, $g(y_{T+1}|\mathbf{y})$, {which is
shown in the appendix to be a proper density function,} represents an
approximation of $p(y_{T+1}|\mathbf{y)}$ that {we refer to as the ABF density.
This density can, in turn, }be estimated via the sequential use of the ABC
draws from $p_{\varepsilon}(\mathbf{\theta|\eta}(\mathbf{y}))$ followed by
draws of $y_{T+1}$ conditional on the draws of $\mathbf{\theta}$.

{Certain} natural questions become immediately relevant: First, what role, if
any, {do the properties} of $p_{\varepsilon}(\mathbf{\theta|\eta(y)})$\textbf{
}play in determining the accuracy of $g(y_{T+1}|\mathbf{y})$\textbf{ }as an
estimate of $p(y_{T+1}|\mathbf{y})$?{ }Second, can we formally characterize
the anticipated loss associated with targeting\textbf{ }$g(y_{T+1}%
|\mathbf{y})$\textbf{ }rather than\textbf{ }$p(y_{T+1}|\mathbf{y})${? }Third,
in practical settings do conclusions drawn regarding $Y_{T+1}$\textbf{ }from
$g(y_{T+1}|\mathbf{y})$\textbf{ }and\textbf{ }$p(y_{T+1}|\mathbf{y})$\textbf{
}differ in any substantial way{? }These questions are tackled sequentially
in\textbf{ }Sections \ref{merg}, \ref{proper} and \ref{numerics 1}
respectively, after a brief review of existing asymptotic results pertaining
to\textbf{ }$p_{\varepsilon}(\mathbf{\theta|\eta}(\mathbf{y}))$ in Section
\ref{asym_abc}.

However, before addressing the above questions, we acknowledge here that the
ABC posterior $p_{\varepsilon}(\mathbf{\theta}|\mathbf{\eta}(\mathbf{y}))$ is
one of several posterior approximations that have been proposed in the
literature. Other such approximations, {for example}, those produced
by\textbf{ }variational Bayes (\citealp{jaakkola2000bayesian};
\citealp{tran2017variational}), {Bayesian synthetic likelihood
(\citealp{BSL2017}), }Bayesian empirical likelihood
{(\citealp{mengersen2013})}, {or bootstrap methods (\citealp{zhu2016})}, could
also be used to construct an approximate predictive. However, to formally
characterize the accuracy of any such approximate predictive, relative to the
exact predictive $p(y_{T+1}|\mathbf{y})$, we must know a good deal about the
theoretical behavior of the posterior approximation itself. This requirement,
and the ensuing regularity of the ABC posterior, partly motivates our focus on
ABC as the inferential approach underpinning the production of an approximate
predictive. In particular, the following section makes substantial use of the
theoretical properties of the ABC posterior in characterizing the accuracy of
ABF relative to exact Bayesian forecasting.

\section{Accuracy of ABF \label{foreast_accuracy}}

It is well-known in the ABC literature that the posterior $p_{\varepsilon
}(\mathbf{\theta|\eta(y)})$ is sometimes a poor approximation to
$p(\mathbf{\theta}|\mathbf{y})$ (\citealp{marinea2012}). What is unknown,
however, is whether or not this same degree of inaccuracy will transfer to the
ABC-based predictive. To this end, we begin by characterizing{ }the difference
between $g(y_{T+1}|\mathbf{y})$ and $p(y_{T+1}|\mathbf{y})$ using the large
sample behavior of $p_{\varepsilon}(\mathbf{\theta|\eta(y)})$ and
$p(\mathbf{\theta}|\mathbf{y})$. In so doing, in Section \ref{merg} we
demonstrate that if both $p_{\varepsilon}(\mathbf{\theta|\eta(y)})$ {and
}$p(\mathbf{\theta}|\mathbf{y})$ are Bayesian consistent for the true value
$\mathbf{\theta}_{0}$, then the densities $g(y_{T+1}|\mathbf{y})$ and
$p(y_{T+1}|\mathbf{y})$ produce the same predictions {asymptotically;} that
is, $g(y_{T+1}|\mathbf{y})$ and $p(y_{T+1}|\mathbf{y})$ `merge' asymptotically
(\citealp{blackwell1962}; \citealp{diaconis1986}). Using the concept of a
proper scoring rule, in Section \ref{proper} we quantify the loss in
forecasting accuracy incurred by using $g(y_{T+1}|\mathbf{y})$ rather than
$p(y_{T+1}|\mathbf{y})$. In Section \ref{numerics 1} we then provide numerical
illustrations of\textbf{ }$g(y_{T+1}|\mathbf{y})$ and $p(y_{T+1}|\mathbf{y})$
for particular models, and for particular choices of summary statistics in the
production of\textbf{ }$g(y_{T+1}|\mathbf{y}).$

We{ first} give a brief overview of certain existing results on the asymptotic
properties of $p_{\varepsilon}(\mathbf{\theta|\eta(y)})$, {which inform} the
{theoretical} results pertaining to approximate forecasting.

\subsection{Asymptotic Properties of ABC posteriors\label{asym_abc}}

We briefly summarize recent theoretical results for ABC as they pertain to our
eventual goal of demonstrating the merging of $g(y_{T+1}|\mathbf{y})$ and
$p(y_{T+1}|\mathbf{y})$. To this end, we draw on the work of \cite{FMRR2016}
but acknowledge here the important contributions by \cite{LF2016b} and
\cite{LF2016a}. As is consistent with the standard approach to {Bayesian
asymptotics }(\citealp{van1998}; \citealp{Ghosh2003}), we view the
conditioning values $\mathbf{y}$ as random and thus, by extension,\textbf{
}$g(y_{T+1}|\mathbf{y})$\textbf{ }and $p(y_{T+1}|\mathbf{y})${. }However, for
ease of notation, we continue to use the lower case notation $\mathbf{y}$ everywhere.

Establishing the asymptotic properties of $p_{\varepsilon}(\mathbf{\theta
|\eta(y)})$ requires simultaneous asymptotics in the tolerance, $\varepsilon$,
and the sample size, $T$. To this end, we denote a hypothetical $T-$dependent
ABC tolerance by $\varepsilon_{T}$. Under relatively weak sufficient
conditions on the prior $p(\mathbf{\theta})$ and the tail behavior of
$\mathbf{\eta}(\mathbf{y})$, plus an identification condition that is
particular to the probability limit of\textbf{ }$\mathbf{\eta}(\mathbf{y})$,
\cite{FMRR2016} prove the {following results }regarding the posterior produced
from the ABC draws in Algorithm 1, {as }$T\rightarrow\infty$: \medskip

\noindent1. The posterior concentrates onto $\mathbf{\theta}_{0}$ {(i.e. is
Bayesian consistent)} for any $\varepsilon_{T}=o(1);$\medskip

\noindent2. The posterior is asymptotically normal for $\varepsilon_{T}%
=o(\nu_{T}^{-1})$, where $\nu_{T}$ is the rate at which the summaries
$\eta(\mathbf{y})$ satisfy a central limit theorem.\medskip

In Section \ref{merg} we show that {under Bayesian consistency, predictions
}generated from $g(y_{T+1}|\mathbf{y})$ will, to all intents and purposes, be
identical to those generated from\textbf{ }$p(y_{T+1}|\mathbf{y})$. {The
asymptotic normality (i.e. }a Bernstein-von Mises result{) in 2. is applied in
Section \ref{proper}. Note that, without making this explicit, we assume that
the tolerance underpinning an ABC posterior is specified in such a way that
the theoretical properties invoked hold. }

\subsection{Merging of Approximate and Exact Predictives\label{merg}}

Let $(\Omega,\mathcal{F},\mathbb{P})$ be a probability space, with
$\mathbb{P}$ a convex class of probability measures on $(\Omega,\mathcal{F})$.
Define a filtration $\{\mathcal{F}_{t}:t\geq0\}$ associated with the
probability space $(\Omega,\mathcal{F},\mathbb{P})$, and let the sequence
$\{y_{t}\}_{t\geq1}$ be adapted to $\{\mathcal{F}_{t}\}$. Define, for
$B\in\mathcal{F}$, the following predictive measures:%
\begin{align*}
P_{\mathbf{y}}(B) &  =\int_{\Omega}\int_{\Theta}p(y_{T+1}|\mathbf{\theta
},\mathbf{y})d\Pi\lbrack\mathbf{\theta}|\mathbf{y}]d\delta_{y_{T+1}}(B),\\
G_{\mathbf{y}}(B) &  =\int_{\Omega}\int_{\Theta}p(y_{T+1}|\mathbf{\theta
},\mathbf{y})d\Pi\lbrack\mathbf{\theta}|\eta(\mathbf{y})]d\delta_{y_{T+1}}(B),
\end{align*}
where\textbf{\ }$\delta_{x}$ denotes the Dirac measure. $P_{\mathbf{y}}%
(\cdot)$ denotes the predictive distribution for the random variable $Y_{T+1}%
$, {conditional on }$\mathbf{y}$, and where parameter uncertainty - integrated
out in the process of producing $P_{\mathbf{y}}(\cdot)~$- is described by the
exact posterior distribution, $\Pi\lbrack\cdot|\mathbf{y}]$, with
density\textbf{\ }$p(\mathbf{\theta}|\mathbf{y})$ (with respect to the
Lebesgue measure). $G_{\mathbf{y}}(\cdot)$ is the ABF predictive and differs
from $P_{\mathbf{y}}(\cdot)$ in its quantification of parameter uncertainty,
which is expressed via $\Pi\lbrack\cdot|\mathbf{\eta(y)}]$ instead of
$\Pi\lbrack\cdot|\mathbf{y}]$, where the former has density\textbf{\ }%
$p_{\varepsilon}(\mathbf{\theta|\eta}(\mathbf{y})).$

The discrepancy between $G_{\mathbf{y}}$ and $P_{\mathbf{y}}$ is entirely due
to the replacement of $\Pi\lbrack\mathbf{\theta}|\mathbf{y}]$ by $\Pi
\lbrack\mathbf{\theta}|\eta(\mathbf{y})]$. In this way, noting that, for any
$B\in\mathcal{F}$,
\[
|G_{\mathbf{y}}(B)-P_{\mathbf{y}}(B)|\leq\int_{\Omega}\int_{\Theta}%
p(y_{T+1}|\mathbf{\theta},\mathbf{y})d\delta_{y_{T+1}}(B)\left\vert
p_{\varepsilon}(\mathbf{\theta|\eta(y)})-p(\mathbf{\theta}|\mathbf{y}%
)\right\vert d\mathbf{\theta}\text{,}%
\]
it is clear that the difference between $G_{\mathbf{y}}$ and $P_{\mathbf{y}}$
is smaller, the smaller is the discrepancy between $p(\mathbf{\theta
}|\mathbf{y})$ and $p_{\varepsilon}(\mathbf{\theta|\eta(y)})$.

Under regularity conditions ({see, for {example,
{\citealp{ghosal1995convergence} or \citealp{ibragimov2013}}) the exact
posterior $p(\mathbf{\theta}|\mathbf{y})$ will concentrate onto
$\mathbf{\theta}_{0}$ as $T\rightarrow\infty.$ {As long as} the relevant
conditions delineated in \cite{FMRR2016} for the Bayesian consistency of
$p_{\varepsilon}(\mathbf{\theta|\eta(y)})$ {are satisfied}, then
$p_{\varepsilon}(\mathbf{\theta|\eta(y)})$ will also concentrate onto
$\mathbf{\theta}_{0}$ as $T\rightarrow\infty$. {Consequently, the discrepancy
between }$p_{\varepsilon}(\mathbf{\theta|\eta(y)})${ and $p(\mathbf{\theta
}|\mathbf{y})$ will disappear in large samples, and mitigate the discrepancy
between }$g(y_{T+1}|\mathbf{y})$ and $p(y_{T+1}|\mathbf{y})${.} The following
theorem formalizes this intuition.}}

\begin{theorem}
\label{thm2} Under Assumption \ref{ass1} in Appendix \ref{thm2_proof}, the
predictive distributions $P_{\mathbf{y}}(\cdot)$ and $G_{\mathbf{y}}(\cdot)$
merge, in the sense that $\rho_{TV}\{P_{\mathbf{y}},G_{\mathbf{y}%
}\}\rightarrow0$ as $T\rightarrow\infty$ and $\varepsilon_{T}\rightarrow0$,
with $\mathbb{P}$-probability 1, where $\rho_{TV}\{P_{\mathbf{y}%
},G_{\mathbf{y}}\}$ denotes the total variation metric: $\sup_{B\in
\mathcal{F}}|P_{\mathbf{y}}(B)-G_{\mathbf{y}}(B)|.$
\end{theorem}

The merging of $P_{\mathbf{y}}$ and $G_{\mathbf{y}}$ is not without precedence
and mimics early results on merging of predictive distributions due to
\cite{blackwell1962}. A connection between merging of predictive distributions
and Bayesian consistency was first discussed in \cite{diaconis1986}, with the
authors viewing Bayesian consistency as implying a \textquotedblleft merging
of inter-subjective opinions\textquotedblright. In their setting, Bayesian
consistency implied that two separate Bayesians with different subjective
prior beliefs would ultimately end up with the same predictive distribution.
(See also \citealp{petrone2014}, for related work).

Our situation is qualitatively different from that considered in
\cite{diaconis1986} in that we are not concerned with Bayesians who have
different \textit{prior} beliefs but Bayesians who are using completely
different means of {assessing} the \textit{posterior} uncertainty about the
parameters $\mathbf{\theta}$. Given the nature of ABC, and the fact that under
suitable conditions posterior concentration can be proven, we have the
{interesting} result that, for a large enough sample, {and under Bayesian
consistency of both }$p_{\varepsilon}(\mathbf{\theta|\eta(y)})${ and
$p(\mathbf{\theta}|\mathbf{y})$, }conditioning inference
about{\ $\mathbf{\theta}$ }on\textbf{\ $\mathbf{\eta}(\mathbf{y})$ }rather
than\textbf{\ $\mathbf{y}$} makes no difference to the probabilistic
statements made about $Y_{T+1}.$ In contexts where inference about
$\mathbf{\theta}$ is simply a building block for Bayesian predictions, and
where sample sizes are sufficiently large, inference undertaken via (posterior
consistent) ABC is sufficient to yield predictions that are virtually
identical to those obtained by an exact (but potentially infeasible or, at the
very least computationally challenging) method.

\subsection{Proper Scoring Rules\label{proper}}

The above merging result demonstrates that in large samples the difference
between $p(y_{T+1}|{\mathbf{y}})$ and $g(y_{T+1}|{\mathbf{y}})$ is likely to
be small. To {formally quantify} the {loss in forecast accuracy} incurred by
using $g(y_{T+1}|\mathbf{y})$ rather than $p(y_{T+1}|\mathbf{y})$, {we use the
concept of a scoring rule}. Heuristically, a scoring rule rewards a forecast
for assigning a high density ordinate (or high probability mass) to the
observed value (so-called `calibration'), often subject to some shape or
`sharpness' criterion. (See \citealp{GR2007} and \citealp{Gea2007} for
expositions). More specifically, we are interested in scoring rules
$S:\mathbb{P}\times\Omega\mapsto\mathbb{R}$ whereby if the forecaster quotes
the predictive distribution $G$ and the value $y$ eventuates, then the reward
(or `score') is $S(G,y).$ {We then define} the \textit{expected} score under
measure $P$\ of the probability forecast $G$, as
\begin{equation}
\mathbb{M}(G,P)=\int_{y\in\Omega}S(G,y)dP(y). \label{m_score}%
\end{equation}
A scoring rule $S(\cdot,\cdot)$ is proper if for all $G,P\in\mathbb{P}$,%
\[
\mathbb{M}(P,P)\geq\mathbb{M}(G,P),
\]
and is strictly proper, relative to $P$, if $\mathbb{M}(P,P)=\mathbb{M}(G,P)$
implies $G=P$. That is, a proper scoring rule is one whereby \textit{if} the
forecasters best judgment is indeed $P$ there is no incentive to quote
anything other than $G=P.$

{Now define the true predictive distribution of the random variable} $Y_{T+1}%
$, {evaluated at} $\theta_{0},$ {as}%
\[
F_{\mathbf{y}}(B)=\int_{\Omega}\int_{\Theta}p(y_{T+1}|\mathbf{\theta
},\mathbf{y})d\delta_{\theta}(\theta_{0})d\delta_{y_{T+1}}(B).
\]
The following result builds on Theorem \ref{thm2} and {presents a theoretical
relationship between }the predictive density functions, $g(y_{T+1}%
|\mathbf{y})$ and $p(y_{T+1}|\mathbf{y})$, in terms of {the expectation of
}proper scoring rules {with respect to }$F_{\mathbf{y}}(\cdot).$\medskip

\begin{theorem}
\label{thm1} Under Assumption \ref{ass1} in Appendix \ref{thm2_proof}, if
$S(\cdot,\cdot)$ is a strictly proper scoring rule,

\begin{enumerate}
\item[(i)] $|\mathbb{M}(P_{\mathbf{y}},F_{\mathbf{y}})-\mathbb{M}%
(G_{\mathbf{y}},F_{\mathbf{y}})|=o_{\mathbb{P}}(1)$;

\item[(ii)] $|\mathbb{E}\left[  \mathbb{M}(P_{\mathbf{y}},F_{\mathbf{y}%
})\right]  -\mathbb{E}\left[  \mathbb{M}(G_{\mathbf{y}},F_{\mathbf{y}%
})\right]  |=o(1)$;

\item[(iii)] {The absolute differences in }(i) and (ii) {are identically zero}
if and only if $\eta(\mathbf{y)}$ is sufficient for $\mathbf{y}$ and
$\varepsilon_{T}=0$.
\end{enumerate}
\end{theorem}

\medskip

The result in (i) {establishes an asymptotic equivalence}{ between }the
expected scores (under{ }$F_{\mathbf{y}}$) of the exact and approximate
predictives, {where the expectation is with respect to }$Y_{T+1}$,
{conditional on }$\mathbf{y}$. {Hence, the result establishes that (under
regularity) as }$T\rightarrow\infty$, {there is no expected loss in accuracy
from basing predictions on an approximation.} The {result} in (ii) is
{marginal of }$\mathbf{y}$ and follows from (i) and the monotonicity property
of integrals. Part (iii) {follows }from the factorization theorem and the
structure of $P_{\mathbf{y}}$ and $G_{\mathbf{y}}$. {All results are, of
course, consistent with the merging result demonstrated earlier, and with
}$P_{\mathbf{y}}$ and $G_{\mathbf{y}}${, by definition, equivalent for any
}$T$ {under sufficiency of }$\eta(\mathbf{y).}$

{If, however}, one {is} willing to make additional assumptions about the
regularity of {$p(\mathbf{\theta}|\mathbf{y})$ and $p_{\varepsilon
}(\mathbf{\theta}|\eta(\mathbf{y}))$}, {one can go }further{ than the result
in Theorem \ref{thm1}, to produce an actual \textit{ranking} of}
$\mathbb{M}(P_{\mathbf{y}},F_{\mathbf{y}})$ and $\mathbb{M}(G_{\mathbf{y}%
},F_{\mathbf{y}})$, which should hold for large $T$ with high probability.
Heuristically, if both $p(\mathbf{\theta}|\mathbf{y})$ and {$p_{\varepsilon
}(\mathbf{\theta}|\eta(\mathbf{y}))$ satisfy a Bernstein-von Mises result
(invoking {Result 2} {in Section \ref{asym_abc}} in the latter case and
standard regularity in the former): for $\phi_{\theta,V}$ a normal density
function with mean $\mathbf{\theta}$ and variance $V$,
\begin{align}
p(y_{T+1}|\mathbf{y}) &  =\int_{\Theta}p(y_{T+1}|\mathbf{\theta}%
,\mathbf{y})\phi_{\widehat{\theta},\mathcal{I}^{-1}}(\mathbf{\theta
})d\mathbf{\theta}+o_{\mathbb{P}}(T^{-1/2})\label{p}\\
g(y_{T+1}|\mathbf{y}) &  =\int_{\Theta}p(y_{T+1}|\mathbf{\theta}%
,\mathbf{y})\phi_{\tilde{\mathbf{\theta}},\mathcal{E}^{-1}}(\mathbf{\theta
})d\mathbf{\theta}+o_{\mathbb{P}}(T^{-1/2}),\label{g}%
\end{align}
where $\hat{\mathbf{\theta}}$ is the maximum likelihood estimator (MLE),
$\mathcal{I}$ is the Fisher information matrix (evaluated at $\mathbf{\theta
}_{0}$), $\tilde{\mathbf{\theta}}$ is the ABC posterior mean and $\mathcal{E}$
is the Fisher information conditional on the statistic $\eta(\mathbf{y})$
(evaluated at $\mathbf{\theta}_{0}$). We assume, {for simplicity}, that
both\textbf{ }$\mathcal{I}^{-1}$ and $\mathcal{E}^{-1}$\textbf{ }are\textbf{
}$O(T^{-1})$, where $\mathcal{I}^{-1}-\mathcal{E}^{-1}$ is {negative
semi-definite}. Now, assuming validity of a second-order Taylor expansion for
$p(y_{T+1}|\mathbf{\theta},\mathbf{y})$\textbf{ }in a neighborhood of\textbf{
}}{$\hat{\mathbf{\theta}}$, }we can expand this function as{
\begin{equation}
p(y_{T+1}|\mathbf{\theta},\mathbf{y})=p(y_{T+1}|\hat{\mathbf{\theta}%
},\mathbf{y})+\left.  \frac{\partial p(y_{T+1}|\mathbf{\theta},\mathbf{y}%
)}{\partial\mathbf{\theta}^{\prime}}\right\vert _{\mathbf{\theta}%
=\hat{\mathbf{\theta}}}(\mathbf{\theta}-\hat{\mathbf{\theta}})+\frac{1}%
{2}(\mathbf{\theta}-\hat{\mathbf{\theta}})^{\prime}\left.  \frac{\partial
^{2}p(y_{T+1}|\mathbf{\theta},\mathbf{y})}{\partial\mathbf{\theta
\partial\theta}^{\prime}}\right\vert _{\mathbf{\theta}=\mathbf{\theta}^{\ast}%
}(\mathbf{\theta}-\hat{\mathbf{\theta}}),\label{ts}%
\end{equation}
for some intermediate value $\mathbf{\theta}^{\ast}.$ {Substituting (\ref{ts})
into (\ref{p}), and recognizing that }$\int_{\Theta}(\mathbf{\theta}%
-\hat{\mathbf{\theta}})\phi_{\widehat{\theta},\mathcal{I}^{-1}}(\mathbf{\theta
})d\mathbf{\theta}$ $=0$, {then yields}}%
\begin{align*}
p(y_{T+1}|\mathbf{y}) &  =\int_{\Theta}p(y_{T+1}|\hat{\mathbf{\theta}%
},\mathbf{y})\phi_{\widehat{\theta},\mathcal{I}^{-1}}(\mathbf{\theta
})d\mathbf{\theta}+\frac{1}{2}\text{tr}\left\{  \left.  \frac{\partial
^{2}p(y_{T+1}|\mathbf{\theta},\mathbf{y})}{\partial\mathbf{\theta
\partial\theta}^{\prime}}\right\vert _{\mathbf{\theta}=\mathbf{\theta}^{\ast}%
}\int_{\Theta}(\mathbf{\theta}-\hat{\mathbf{\theta}})(\mathbf{\theta}%
-\hat{\mathbf{\theta}})^{\prime}\phi_{\widehat{\theta},\mathcal{I}^{-1}%
}(\mathbf{\theta})d\mathbf{\theta}\right\}  \\
&  +o_{\mathbb{P}}(T^{-1/2})\\
&  =p(y_{T+1}|\hat{\mathbf{\theta}},\mathbf{y})+O_{\mathbb{P}}(1)O(T^{-1}%
)+o_{\mathbb{P}}(T^{-1/2})\\
&  =p(y_{T+1}|\hat{\mathbf{\theta}},\mathbf{y})+o_{\mathbb{P}}(1).
\end{align*}
Similarly, we have for $g(y_{T+1}|\mathbf{y})$ {in (\ref{g}):}
\[
g(y_{T+1}|\mathbf{y})=p(y_{T+1}|\tilde{\mathbf{\theta}},\mathbf{y}%
)+o_{\mathbb{P}}(1).
\]

{Heuristically, for large }$T$, {under the approximate} Gaussianity of
${\hat{\theta}}$ and ${\tilde{\theta}}$, {we can view}
$p(y_{T+1}|\hat{\mathbf{\theta}},\mathbf{y})-p(y_{T+1}|{\mathbf{\theta}_{0}%
},\mathbf{y})$ and $p(y_{T+1}|\tilde{\mathbf{\theta}},\mathbf{y})-$
$p(y_{T+1}|{\mathbf{\theta}_{0}},\mathbf{y})$ as {approximately Gaussian} with
mean $0$, {but with the former having a smaller variance than the latter
}({even though these un-normalized quantities have variances that are both
collapsing to zero as}{ }$T\rightarrow\infty$){.} Therefore, {on average},
{the error }$p(y_{T+1}|\hat{\mathbf{\theta}},\mathbf{y})-p(y_{T+1}%
|{\mathbf{\theta}_{0}},\mathbf{y})$, should be {smaller than the error
}$p(y_{T+1}|\tilde{\mathbf{\theta}},\mathbf{y})-p(y_{T+1}|{\mathbf{\theta}%
_{0}},\mathbf{y})$, so that, for $S(\cdot,\cdot)$ a proper scoring rule, on
average,
\begin{align}
\int_{\Omega}S(p(y_{T+1}|{\mathbf{\theta}}_{0},\mathbf{y}),y_{T+1}%
)p(y_{T+1}|{\mathbf{\theta}}_{0},\mathbf{y})dy_{T+1} &  \geq\int_{\Omega
}S(p(y_{T+1}|\hat{\mathbf{\theta}},\mathbf{y}),y_{T+1})p(y_{T+1}%
|{\mathbf{\theta}}_{0},\mathbf{y})dy_{T+1}\nonumber\\
&  \geq\int_{\Omega}S(p(y_{T+1}|\tilde{\mathbf{\theta}},\mathbf{y}%
),y_{T+1})p(y_{T+1}|{\mathbf{\theta}}_{0},\mathbf{y})dy_{T+1}%
.\label{score_inequal}%
\end{align}
That is, using the notation defined in (\ref{m_score}),\textbf{ }one would
expect that, for large enough $T,$
\begin{equation}
\mathbb{M}(P_{\mathbf{y}},F_{\mathbf{y}})\geq\mathbb{M}(G_{\mathbf{y}%
},F_{\mathbf{y}}),\label{m_inequal}%
\end{equation}
{and - as accords with intuition -} {predictive accuracy to be greater when
based on the exact predictive distribution.\footnote{We {reiterate that the
derivation of the result in (\ref{m_inequal}) is based on asymptotic
approximations of the \textit{unscaled} quantities}, $p(y_{T+1}|\mathbf{y})$
and $g(y_{T+1}|\mathbf{y})$, which (in common with all asymptotic results
pertaining to unnormalized quantities) is valid for large but finite $T.$}}

In practice of course, in a situation in which exact inference is deemed to be
infeasible, measurement of this loss is also infeasible, since $p(y_{T+1}%
|\mathbf{y})$ is inaccessible. However, it is of interest - in experimental
settings, in which both $g(y_{T+1}|\mathbf{y})$ and $p(y_{T+1}|\mathbf{y})$
\textit{can} be computed - to gauge the extent of this discrepancy, in
particular for different choices of $\mathbf{\eta}(\mathbf{y}).$ This then
gives us some insight into what might be expected in the more realistic
scenario in which the exact predictive cannot be computed and the ABF density
is the only option. Furthermore, even in situations in which $p(y_{T+1}%
|\mathbf{y})$\textit{\ }can\textit{ }be accessed, but only via a bespoke,
finely-tuned MCMC algorithm, a finding that the approximate predictive
produced via the simpler, more readily automated and less computationally
burdensome ABC algorithm, is very similar to the exact, is consequential for
practitioners. {We pursue such matters in the following {Section
\ref{numerics 1}}, }{with the specific matter of asymptotic merging - and the
role played therein by Bayesian consistency - treated in Section
\ref{numerics_consistency}. }

\subsection{Numerical Illustrations\label{numerics 1}}

\subsubsection{Example: Integer Autoregressive Model\label{inar_num}}

We begin by illustrating the approximate forecasting methodology for the case
of a discrete random variable, in which case the object of interest is a
predictive mass function. To do so, we adopt an integer autoregressive model
of order one (INAR(1)) as the data generating process. The INAR(1) model is
given as
\begin{equation}
y_{t}=\rho\circ y_{t-1}+\varepsilon_{t},\label{inar}%
\end{equation}
where $\circ$ is the binomial thinning operator defined as
\begin{equation}
\rho\circ y_{t-1}=\sum_{j=0}^{y_{t-1}}B_{j}(\rho),\label{bt}%
\end{equation}
and where $B_{1}(\rho)$, $B_{2}(\rho)$,..., $B_{y_{t-1}}(\rho)$ are $i$%
.$i$.$d$. {Bernoulli random variables each with}
\[
\Pr(B_{j}(\rho)=1)=1-\Pr(B_{j}(\rho)=0)=\rho.
\]
In the numerical {illustration} we take $\varepsilon_{t}$ to be $i$.$i$.$d$.
Poisson with intensity parameter $\lambda$.

The INAR(1) model sits within the broader class of integer-valued ARMA
(INARMA) models, which has played a large role in the modeling and forecasting
of count time series data. See \cite{jung2006} for a review, and
\cite{drost2009efficient} and \cite{McCabe2011} for contributions. Of
particular note is the work by \cite{martin2014}, in which the INARMA model is
estimated `indirectly' via efficient method of moments
(\citealp{GallantTauchen96}), which is similar in spirit to ABC. No
investigation of forecasting under this `approximate' inferential paradigm is
however undertaken.

Relevant also is the work of \cite{neal2007} in which an MCMC scheme for the
INARMA class is devised, and from which an exact predictive could be
estimated. However, given the very simple parameterization of (\ref{inar}), we
evaluate the exact posterior for $\mathbf{\theta}=(\rho,\lambda)^{\prime}$
numerically using deterministic integration, and estimate the exact predictive
in (\ref{exact_predictive}) by taking a simple weighted average of the
ordinates of the one-step-ahead conditional predictive associated with the
model. Given the structure of (\ref{inar}) {this conditional} predictive mass
function is defined by the convolution of the two unobserved random variables,
$\rho\circ y_{T}$ and $\varepsilon_{T},$ as\textbf{ }%
\begin{equation}
\Pr\left(  {{Y_{T+1}=y}}_{T+1}{{|}\mathbf{y},}\theta\right)  =\sum
\limits_{s=0}^{\min\{y_{T+1},y_{T}\}}\Pr\left(  B_{y_{T}}^{\rho}=s\right)
{\Pr(\varepsilon}_{T+1}=y{_{T+1}-s),}\label{transition}%
\end{equation}
where $\Pr\left(  B_{y_{T}}^{\rho}=s\right)  $ denotes the probability that a
binomial random variable associated with $y_{T}$ replications (and a
probability of `success', $\rho$, on each replication) {takes a value of $s$,
and where} $\Pr(\varepsilon_{T+1}=y_{T+1}-s)$ denotes the probability that a
Poisson random variable takes a value of\textbf{ }$y_{T+1}-s.$

We generate a sample of size $T=100$ from the model in (\ref{inar}) and
(\ref{bt}), with\textbf{ }$\mathbf{\theta}_{0}=(\rho_{0},\lambda_{0})^{\prime
}=(0.4,2)^{\prime}$. Prior information on $\mathbf{\theta}$ is specified as
$U[0,1]\times U[0,10]$.\footnote{In this and the following sections we use the
simplest possible priors, including truncated uniform priors on location
parameters. We acknowledge that these prior choices will have some influence
on the posterior densities produced, both exact and approximate. However,
given that the sample sizes are reasonable (and large in some cases) we do not
expect that influence to be substantial, nor for the conclusions regarding
predictive performance to qualitatively alter. In particular we emphasize that
the \textit{same} priors are used to generate both the exact and approximate
posteriors in all cases.} We implement ABC using a nearest-neighbour version
of Algorithm 1. This version of ABC replaces Step-3 in Algorithm \ref{ABC}
with the following selection step:\smallskip

\noindent\textbf{3.} Select all $\theta^{i}$ associated with the
$\alpha=\delta/N$ smallest distances $d_{2}\{\eta(\mathbf{z}^{i}%
),\eta(\mathbf{y})\}$ for some $\delta$. \smallskip

\begin{figure}[ptb]
\centering
\setlength\figureheight{4.0cm}
\setlength\figurewidth{4.25cm}
\input{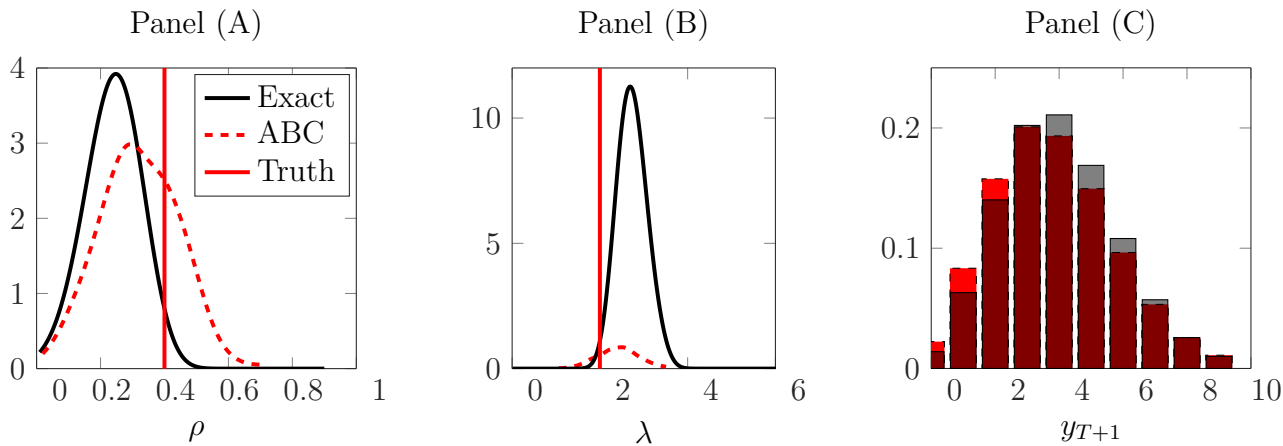}\caption{Panels (A) and (B) depict the marginal
posteriors (exact and ABC) for $\rho$ and $\lambda$, respectively. The red
vertical line (denoted by `Truth' in the key) represents the true value of the
relevant parameter in {both panels. }Panel (C) plots the one-step-ahead
predictive {mass} functions - both exact and approximate (ABC-based). {The
brown shading corresponds to an overlap of exact and approximate predictive
probabilities. The red shading indicates when the approximate probabilities
exceed the exact, with the grey shading indicating the reverse situation.}}%
\label{fig0}%
\end{figure}

For this experiment, the nearest-neighbour version of ABC is implemented by
retaining the simulated draws that lead to the smallest $\alpha=0.01$ of the
$N=20,000$ simulated draws based on a single vector of summary statistics
comprising the sample mean of $\mathbf{y}$, denoted as $\bar{y}$, and the
first three sample autocovariances,\textbf{ }$\gamma_{l}=cov(y_{t},y_{t-l})$,
$l=1,2,3$: $\eta(\mathbf{y})=(\bar{y},\gamma_{1},\gamma_{2},\gamma
_{3})^{\prime}$. Given the latent structure of (\ref{bt}) no reduction to
sufficiency occurs; hence neither this, nor any other set of summaries will
replicate the information in\textbf{ }$\mathbf{y}$, and $p_{\varepsilon
}(\mathbf{\theta|\eta(y)})$ will thus be distinct from $p(\mathbf{\theta
}|\mathbf{y})$.\textbf{ }{As is evident by the plots in Panels A and B of
Figure \ref{fig0}, the exact and ABC posteriors for each element of }$\theta$
{are indeed quite different one from the other. In contrast, in Panel C the
exact and approximate predictive mass functions (with the latter estimated by
taking the average of the conditional predictives in (\ref{transition}) over
the ABC draws of }$\theta$) are seen to be an extremely close
match.\footnote{We note, with reference to the marginal posteriors of
$\lambda$, that the ABC posterior places much more mass over the entire prior
support for $\lambda$, given as $[0,10]$, than does the exact posterior; hence
the very marked difference in their shapes.} 

To illustrate the results of Theorem \ref{thm1}, we construct a series of 100
expanding window one-step-ahead predictive distributions (beginning with a
sample size of $T=100$), and report the average (over 100 one-step-ahead
predictions) of the log score (LS) and quadratic score (QS) in Table
\ref{tab:INARscores}, using the `observed' value of\textbf{ }$y_{T+1}$ that is
also simulated.\footnote{In an expanding window one-step-ahead prediction
scheme, the initial sample, say from period $1$ to $T$, is used to produce a
one-step-ahead prediction for period $T+1$. At the next iteration, we use
observations up to, and including, time $T+1$ to produce a prediction for
period $T+2$. This expanding window procedure then iterates until some
pre-specified period, say $T+K$.} (See \citealp{Gea2007}, for details of these
particular scoring rules.) The assumptions under which Theorem 2 hold can be
demonstrated analytically in this case, including the Bayesian consistency of
$p_{\varepsilon}(\mathbf{\theta|\eta(y)})$; {see Appendix A.4.1.} It is
immediately obvious that, at least according to these two scoring rules, and
to two decimal places, the predictive accuracy of $g(y_{T+1}|\mathbf{y})$ and
$p(y_{T+1}|\mathbf{y})$ is equivalent, even for this relatively small sample
size.\footnote{Additional simulation results, not reported for brevity,
demonstrate that the qualitative nature of this result is not sensitive to the
choice of $\mathbf{\theta}_{0}$.}

In addition, it is important to note that the computational time required to
produce the exact predictive, via rectangular integration over the prior grid,
is just under four and a half minutes, which is approximately 18 times greater
than the time required to construct the approximate predictive via ABC.
Therefore, in this simple example, we see that ABF offers a substantial speed
improvement over the exact predictive, {with no loss} in predictive
accuracy.\footnote{Given the independent nature of ABC sampling, we are able
to exploit parallel computing. This is done using the standard `parfor'
function in MATLAB. All computations are conducted on an Intel Xeon E5-2630
2.30GHz dual {processor} ({each processor with}{ 6 cores}) with 16GB RAM.
{Note that all computation times quoted in the paper are `time elapsed' or
`wall-clock' time.}}

\begin{table}[h]
\caption{{Log score} (LS) and quadratic {score} (QS) {associated with the}
approximate predictive $g(y_{T+1}|\mathbf{y})$, and the exact {predictive,}
$p(y_{T+1}|\mathbf{y})$, each computed as {an} {average} over a series of
(expanding window) {100} one-step-ahead predictions. The predictive with
highest average score is in bold.\bigskip}%
\label{tab:INARscores}%
\centering%
\begin{tabular}
[c]{rrr}\hline\hline
& ABF & Exact\\
LS & \textbf{-1.89} & \textbf{-1.89}\\
QS & \textbf{0.17} & \textbf{0.17}\\\hline
\end{tabular}
\end{table}

We do emphasize at this point that refinements of Algorithm 1 based on either
post-sampling corrections (\citealp{Beaumont2025}; \citealp{Blum2010}), or the
insertion of MCMC or sequential Monte Carlo steps
(\citealp{marjoram2003markov}; \citealp{Sisson2007};
\citealp{beaumont2009adaptive}) may well improve the accuracy with which the
exact posteriors are approximated. However, the key message - both here and in
what follows - is that a poor match between exact and approximate posteriors
does not necessarily translate into a corresponding poor match at the
predictive level; hence, we choose to use the simplest form of the algorithm
in all illustrations.

\subsubsection{Example: Moving Average Model\label{ma}}

We now explore an example from the canonical class of time series models for a
\textit{continuous}\textbf{ }random variable, namely the Gaussian
autoregressive moving average (ARMA) class. We simulate $T=500$ observations
from an invertible moving average model of order 2 (MA(2)),
\begin{equation}
y_{t}=\varepsilon_{t}+\theta_{1}\varepsilon_{t-1}+\theta_{2}\varepsilon
_{t-2},\label{ma2}%
\end{equation}
where $\varepsilon_{t}\sim i.i.d.N(0,\sigma^{2})$, and the true values of the
unknown parameters are given by $\theta_{10}=0.8$, $\theta_{20}=0.6$ and
$\sigma_{0}=1.0$.\footnote{Similar to the INAR example, additional simulation
results in this MA(2) example, not reported for brevity, demonstrate that the
qualitative results are not sensitive to the choice of $\mathbf{\theta}_{0}$.}
{We specify the following priors: $\theta_{1}\sim U(0,0.99)$, $\theta_{2}\sim
U(0,0.99)$ and $\sigma\sim U(0.1,3)$.} Inference on $\mathbf{\theta}%
=(\theta_{1},\theta_{2},\sigma)^{\prime}$ is conducted via ABC using the
sample autocovariances as summary statistics, with $\eta^{(l)}(\mathbf{y}%
)=(\gamma_{0},\gamma_{1},...,\gamma_{l})^{\prime}$, and $\gamma_{l}%
=cov(y_{t},y_{t-l})$. Four alternative sets of $\eta^{(l)}(\mathbf{y})$ are
considered in this case, with $l=1,2,3,4$. {The} {one-step-ahead} predictive
distributions $g^{(l)}(y_{T+1}|\mathbf{y})$ {are estimated} for each set by
using the selected draws, $\mathbf{\theta}^{i}$, $i=1,2,...,N,$ (again, via a
nearest-neighbour version of Algorithm 1) from $p_{\varepsilon}(\mathbf{\theta
}|\eta^{(l)}(\mathbf{y}))$ to define $p(y_{T+1}|\mathbf{\theta}^{i}%
,\mathbf{y})$, from which draws $y_{T+1}^{i}$, $i=1,2,...,N,$ are taken and
used to produce a kernel density estimate of $g^{(l)}(y_{T+1}|\mathbf{y})$. We
note that the moving average dependence in (\ref{ma2}) means that reduction to
a sufficient set of statistics of dimension smaller than $T$\ is not feasible.
Hence, none of the sets of statistics considered here are sufficient for
$\mathbf{\theta}$ and\textbf{\ }$p_{\varepsilon}(\mathbf{\theta}|\eta
^{(l)}(\mathbf{y}))$ is, once again,\textbf{ }distinct from $p(\mathbf{\theta
}|\mathbf{y})$ for all $l.$

\begin{figure}[h]
\centering
\setlength\figureheight{3.75cm}
\setlength\figurewidth{7cm}
\input{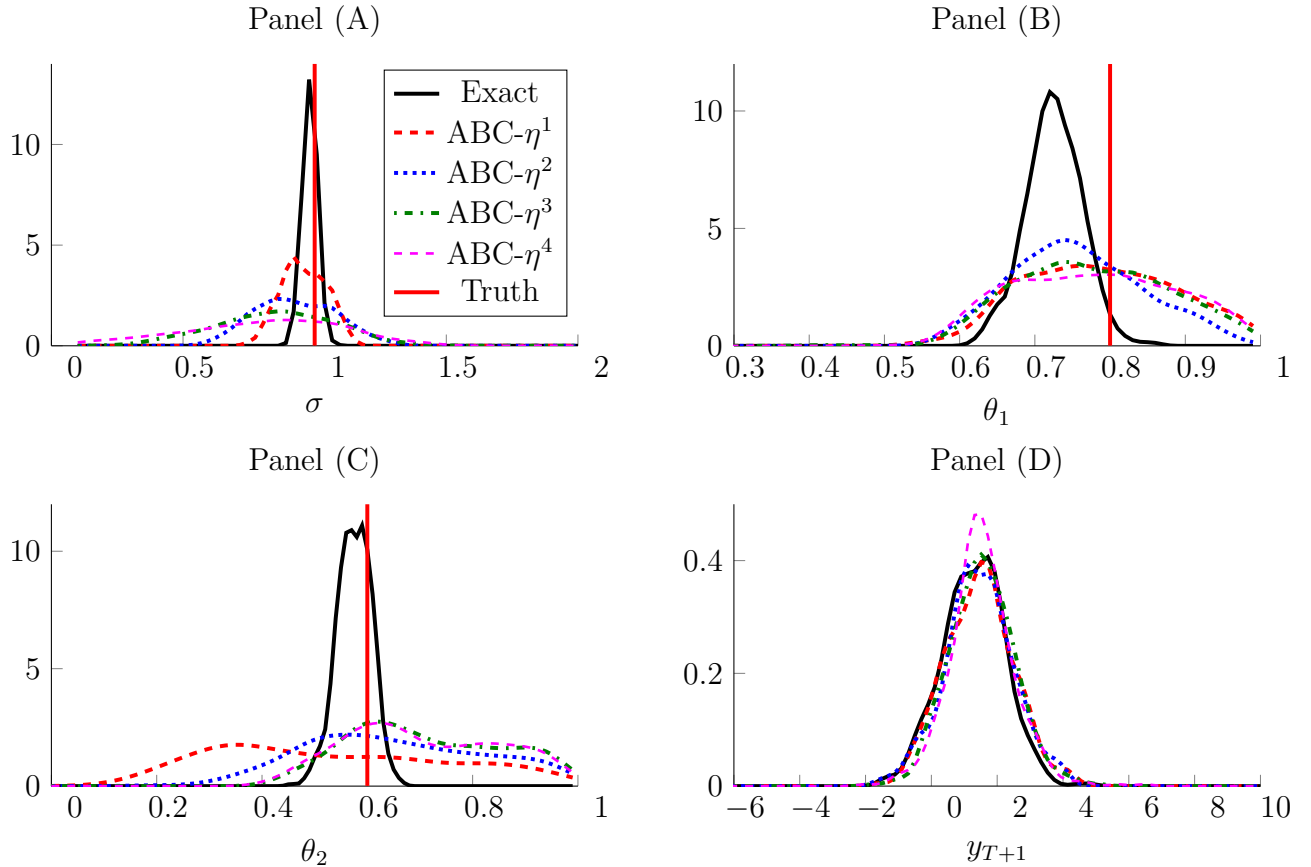}\caption{Panels (A), (B) and (C) depict the marginal
posteriors (exact and ABC) for $\sigma$, $\theta_{1}$ and $\theta_{2}$
respectively. {The four approximate posteriors are based on the sets of
summaries indicated in the key included in Panel A}. Panel (D) plots the
one-step-ahead predictive densities - both exact and approximate (ABC-based).
The red vertical line (denoted by `Truth' in the key) represents the true
value of the relevant parameter in Panels (A), (B) and (C).}%
\label{fig1}%
\end{figure}

Panels (A)-(C) in Figure \ref{fig1} depict the marginal posteriors for each of
the three parameters: the four ABC posteriors are given by the dotted and
dashed curves of various types, with the relevant summary statistic (vector)
indicated in the key appearing in Panel A. The exact marginals (the full
curves) for all parameters are computed using the sparse matrix representation
of the MA(2) process in an MCMC algorithm comprised of standard
Gibbs-Metropolis-Hastings (MH) steps (see, in particular,
\citealp{chan2013moving}). All five densities are computed using\textbf{\ }%
$500$ draws of the relevant parameter. For the ABC densities this is achieved
by retaining (approximately) the smallest 0.5\% of the\textbf{\ }distances in
Algorithm 1, based on $N=111,803$ total draws.\footnote{An explanation of this
particular choice for the selected proportion (and, hence, $N$) is provided in
the next section.} For the exact posterior this is achieved by running the
chain for $N=20,000$ iterates (after a burn-in of 5000) and selecting every
$40^{th}$ draw.

Panel (D) of Figure \ref{fig1} plots the one-step-ahead predictive densities -
both approximate and exact. As is consistent with the previous example, the
contrast between the two sets of graphs in Figure \ref{fig1} is stark. The ABC
posteriors in Panels (A)-(C) are all very inaccurate representations of the
corresponding exact marginals, in addition to being, in some cases, very
different one from the other. In contrast, in Panel D three of the four ABF
predictives (associated with $\eta^{(1)}(\mathbf{y})$, $\eta^{(2)}%
(\mathbf{y})$ and $\eta^{(3)}(\mathbf{y})$) are all very similar, one to the
other, and \textit{extremely accurate} as representations of the exact
predictive; indeed, the approximate predictive generated by $\eta
^{(4)}(\mathbf{y})$ is also relatively close to all other densities.

We now numerically illustrate the content of Section \ref{proper}, by
performing a similar exercise to that undertaken in the previous
section:\textbf{ }we\textbf{ }construct a series of 500 expanding window
one-step-ahead predictive distributions (beginning with a sample size of
$T=500$) and record the average LS, QS and cumulative rank probability score
(CRPS) for each case in Table \ref{tab:MA2scores}. It is clear that the
MCMC-based predictive, which serves as a simulation-based estimate of
$p(y_{T+1}|\mathbf{y})$, generates the highest average score, as is consistent
with (\ref{m_inequal}). Nevertheless, the ABF predictives yield average scores
that are \textit{nearly identical} to those based on MCMC, indeed in one case
(for $l=2$) equivalent to two decimal places. That is, the extent of the loss
associated with the use of insufficient summaries is absolutely minimal.
Moreover, we note that the computational time required to produce the
MCMC-based estimate of the exact predictive {for the case of $T=500$ is just
over 6 minutes, which i}s approximately 115 tim{es }greater than that required
to produce any of the approximate predictives. In any real-time exercise in
which repeated production of such predictions were required, the \textit{vast}
speed improvement yielded by ABF in this example, and with such minimal loss
of accuracy, could be of enormous practical benefit.

\begin{table}[h]
\caption{{Log score }(LS), quadratic {score} (QS) and cumulative rank
probability {score} (CRPS) {associated with the} approximate predictive
{density} $g^{(l)}(y_{T+1}|\mathbf{y})$, $l=1,2,3,4,$ and the exact
{MCMC-based predictive,} $p(y_{T+1}|\mathbf{y})$, each computed as {an}
{average} over a series of 500\textbf{ }(expanding window) one-step-ahead
predictions. The predictive with highest average score is in bold.\bigskip}%
\label{tab:MA2scores}%
\centering
\begin{tabular}
[c]{rlrrrr}\hline\hline
& \multicolumn{1}{l}{$l=1$} & \multicolumn{1}{l}{$l=2$} &
\multicolumn{1}{l}{$l=3$} & \multicolumn{1}{l}{$l=4$} &
\multicolumn{1}{l}{MCMC}\\\hline
LS & -1.43 & -1.42 & -1.43 & -1.43 & \textbf{-1.40}\\
QS & 0.28 & 0.28 & 0.28 & 0.28 & \textbf{0.29}\\
CRPS & -0.57 & \textbf{-0.56} & -0.57 & -0.57 & \textbf{-0.56}\\\hline
\end{tabular}
\end{table}

\subsubsection{Numerical evidence of merging\label{numerics_consistency}}

{In this final sub-section we illustrate the matter of predictive merging and
posterior consistency. To this end, we now }consider data $\mathbf{y}$
simulated from (\ref{ma2}), using\textbf{\ }increasing sample sizes: $T=500$,
$T=2000$, $T=4000$ {and $T=5000$}. {We {also }now make explicit that, of the
four sets of summaries that we continue to use in the illustration,\ the three
sets, }$\eta^{(2)}(\mathbf{y}),$ $\eta^{(3)}(\mathbf{y})$ and $\eta
^{(4)}(\mathbf{y})$ are such that $p_{\varepsilon}(\mathbf{\theta}|\eta
^{(l)}(\mathbf{y}))\,$is Bayesian consistent (see Appendix A.4.2 for this
demonstration), {whilst }$\eta^{(1)}(\mathbf{y})$ can be readily shown to
\textit{not} satisfy the sufficient conditions that guarantee Bayesian consistency.

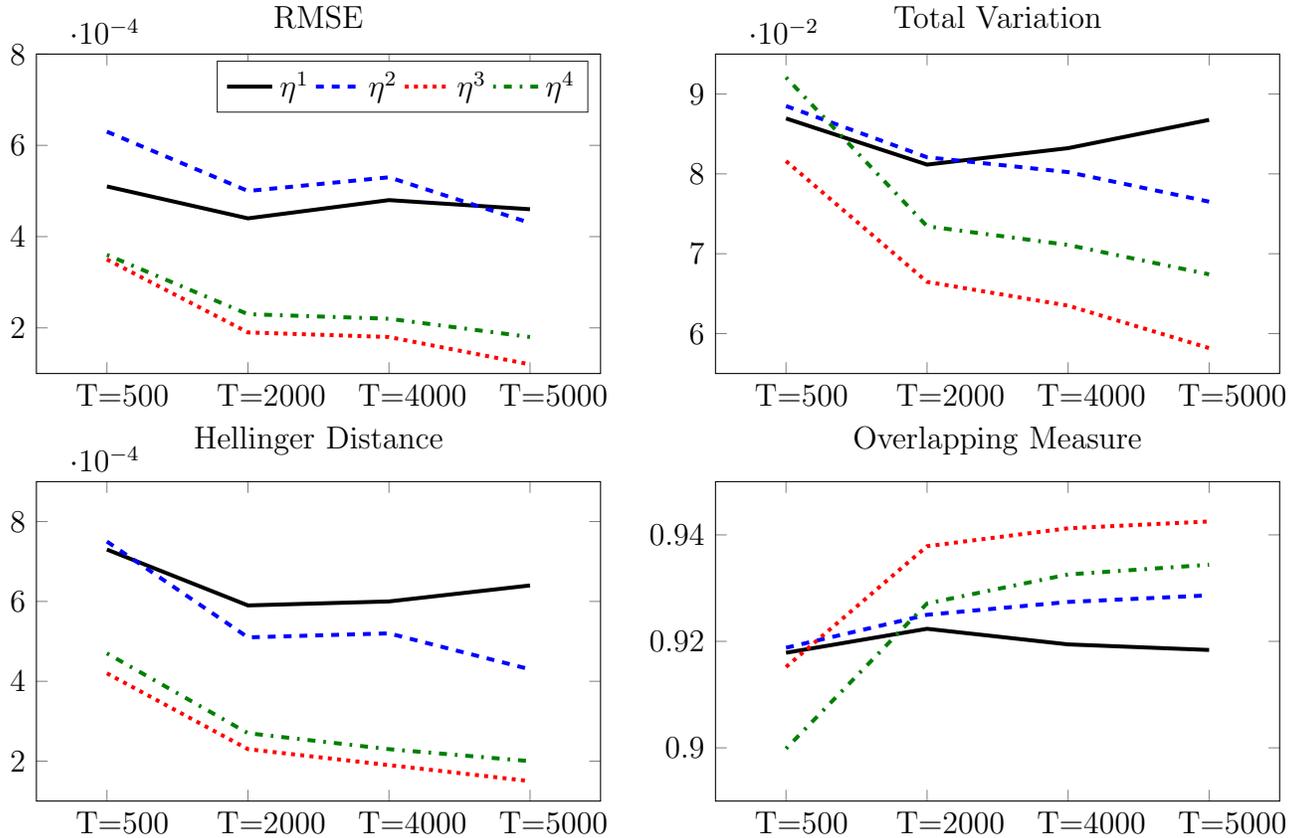
\begin{figure}[ptb]
\centering
\setlength\figureheight{4.25cm}
\setlength\figurewidth{7.5cm}
%
%
%
%

\definecolor{mycolor1}{rgb}{0,0.498039215803146,0}

\begin{tikzpicture}

\begin{axis}[%
width=\figurewidth,
height=\figureheight,
scale only axis,
xmin=0.5,
xmax=4.5,
xtick={1,2,3,4},
xticklabels={T=500,T=2000,T=4000,T=5000},
ymin=0.0001,
ymax=0.0008,
name=plot1,
title={RMSE},
legend style={at={(0.319587185094746,0.8166952881917019)},anchor=south west,legend columns=4,draw=darkgray!60!black,fill=white,legend cell align=left}
]
\addplot [
color=black,
solid,
line width=1.5pt
]
table[row sep=crcr]{
1 0.00051\\
2 0.00044\\
3 0.00048\\
4 0.00046\\
};
\addlegendentry{$\eta^{1}$};

\addplot [
color=blue,
dashed,
line width=1.5pt
]
table[row sep=crcr]{
1 0.00063\\
2 0.0005\\
3 0.00053\\
4 0.00043\\
};
\addlegendentry{$\eta^{2}$};

\addplot [
color=red,
dotted,
line width=1.5pt
]
table[row sep=crcr]{
1 0.00035\\
2 0.00019\\
3 0.00018\\
4 0.00012\\
};
\addlegendentry{$\eta^{3}$};

\addplot [
color=mycolor1,
dash pattern=on 1pt off 3pt on 3pt off 3pt,
line width=1.5pt
]
table[row sep=crcr]{
1 0.00036\\
2 0.00023\\
3 0.00022\\
4 0.00018\\
};
\addlegendentry{$\eta^{4}$};

\end{axis}

\begin{axis}[%
width=\figurewidth,
height=\figureheight,
scale only axis,
xmin=0.5,
xmax=4.5,
xtick={1,2,3,4},
xticklabels={T=500,T=2000,T=4000,T=5000},
ymin=0.055,
ymax=0.095,
name=plot2,
at=(plot1.right of south east),
anchor=left of south west,
title={Total Variation}
]
\addplot [
color=black,
solid,
line width=1.5pt,
forget plot
]
table[row sep=crcr]{
1 0.08694\\
2 0.08116\\
3 0.08322\\
4 0.08676\\
};
\addplot [
color=blue,
dashed,
line width=1.5pt,
forget plot
]
table[row sep=crcr]{
1 0.0885\\
2 0.08208\\
3 0.08022\\
4 0.07652\\
};
\addplot [
color=red,
dotted,
line width=1.5pt,
forget plot
]
table[row sep=crcr]{
1 0.0816\\
2 0.06648\\
3 0.0635\\
4 0.05818\\
};
\addplot [
color=mycolor1,
dash pattern=on 1pt off 3pt on 3pt off 3pt,
line width=1.5pt,
forget plot
]
table[row sep=crcr]{
1 0.09208\\
2 0.07344\\
3 0.0711\\
4 0.06742\\
};
\end{axis}

\begin{axis}[%
width=\figurewidth,
height=\figureheight,
scale only axis,
xmin=0.5,
xmax=4.5,
xtick={1,2,3,4},
xticklabels={T=500,T=2000,T=4000,T=5000},
ymin=0.89,
ymax=0.95,
name=plot4,
at=(plot2.below south west),
anchor=above north west,
title={Overlapping Measure}
]
\addplot [
color=black,
solid,
line width=1.5pt,
forget plot
]
table[row sep=crcr]{
1 0.91786\\
2 0.92236\\
3 0.91943\\
4 0.918381374\\
};
\addplot [
color=blue,
dashed,
line width=1.5pt,
forget plot
]
table[row sep=crcr]{
1 0.9188\\
2 0.92499\\
3 0.9274\\
4 0.928642021\\
};
\addplot [
color=red,
dotted,
line width=1.5pt,
forget plot
]
table[row sep=crcr]{
1 0.91519\\
2 0.93787\\
3 0.94124\\
4 0.942521937\\
};
\addplot [
color=mycolor1,
dash pattern=on 1pt off 3pt on 3pt off 3pt,
line width=1.5pt,
forget plot
]
table[row sep=crcr]{
1 0.89984\\
2 0.92712\\
3 0.93257\\
4 0.934394054\\
};
\end{axis}

\begin{axis}[%
width=\figurewidth,
height=\figureheight,
scale only axis,
xmin=0.5,
xmax=4.5,
xtick={1,2,3,4},
xticklabels={T=500,T=2000,T=4000,T=5000},
ymin=0.0001,
ymax=0.0009,
at=(plot4.left of south west),
anchor=right of south east,
title={Hellinger Distance}
]
\addplot [
color=black,
solid,
line width=1.5pt,
forget plot
]
table[row sep=crcr]{
1 0.00073\\
2 0.00059\\
3 0.0006\\
4 0.00064\\
};
\addplot [
color=blue,
dashed,
line width=1.5pt,
forget plot
]
table[row sep=crcr]{
1 0.00075\\
2 0.00051\\
3 0.00052\\
4 0.00043\\
};
\addplot [
color=red,
dotted,
line width=1.5pt,
forget plot
]
table[row sep=crcr]{
1 0.00042\\
2 0.00023\\
3 0.00019\\
4 0.00015\\
};
\addplot [
color=mycolor1,
dash pattern=on 1pt off 3pt on 3pt off 3pt,
line width=1.5pt,
forget plot
]
table[row sep=crcr]{
1 0.00047\\
2 0.00027\\
3 0.00023\\
4 0.0002\\
};
\end{axis}

\end{tikzpicture}%
\caption{The four panels depict numerical
approximations to the measures in \eqref{rmse}-\eqref{ovl}. The key in the
upper-left-hand panel indicates the set of summaries that underpins the
ABC-based predictive used in each sequence of computations over $T.$}%
\label{statplot}%
\end{figure}

We document the merging across four separate {measures}; with all results
represented as averages over 100 synthetic samples. {We compute the RMSE}
based on the distance between the {CDF} for the approximate and exact
predictives, {as a numerical approximation of}
\begin{equation}
\int\left(  dP_{\mathbf{y}}-dG_{\mathbf{y}}\right)  ^{2}d\mu, \label{rmse}%
\end{equation}
for $\mu$ the Lebesgue measure. Similarly, {we compute (numerical
approximations of) the total variation metric},
\begin{equation}
\rho_{TV}\{P_{\mathbf{y}},G_{\mathbf{y}}\}=\sup_{B\in\mathcal{F}%
}|P_{\mathbf{y}}(B)-G_{\mathbf{y}}(B)|, \label{tv}%
\end{equation}
the {Hellinger distance},%
\begin{equation}
\rho_{H}\{P_{\mathbf{y}},G_{\mathbf{y}}\}=\left\{  \frac{1}{2}\int\left[
\sqrt{{dP}_{\mathbf{y}}}-\sqrt{{dG}_{\mathbf{y}}}\right]  ^{2}d\mu\right\}
^{1/2}, \label{hel}%
\end{equation}
{and the overlapping measure (OVL) (see \citealp{blomstedt2015}) defined as,}
\begin{equation}
\left[  \int\min\{p(y_{T+1}|\mathbf{y}),g(y_{T+1}|\mathbf{y})\}dy_{T+1}%
\right]  ^{2}. \label{ovl}%
\end{equation}
Small RMSE, supremum and Hellinger distances indicate closeness of the
approximate and exact predictive distributions, while {large values} of OVL
indicate a {large degree} of overlap between the two distributions. These four
measures are presented graphically in Figure \ref{statplot}.

All four panels in Figure \ref{statplot} illustrate precisely the role played
by Bayesian consistency in producing a merging of predictive distributions,
{in accordance with }Theorem \ref{thm2}. Specifically, the RMSE, total
variation and Hellinger distances uniformly decrease, while the OVL measure
uniformly increases, as $T$ increases, for the cases of ABF conducted with
$\eta^{(l)}(\mathbf{y})$ for $l=2,3,4$ (all of which are associated with
Bayesian consistent inference). Only in the case of ABF based on $\eta
^{(1)}(\mathbf{y})$ (for which $p_{\varepsilon}(\mathbf{\theta}|\eta
^{(1)}(\mathbf{y}))$ is not Bayesian consistent) is a uniform decline for RMSE
and the total variation and Hellinger distances, not in evidence, and a
uniform increase in OVL not observed.

{We }{comment here} that in order to satisfy the theoretical results discussed
in Section \ref{asym_abc}, we require that the number of draws taken for the
ABC algorithm increases with $T$. {This is a consequence of replacing} the
acceptance step in Algorithm 1 by a nearest-neighbour selection step, with
draws of $\mathbf{\theta}$ being retained only if they fall below a certain
left-hand-tail quantile of the simulated distances. The theoretical results in
\cite{FMRR2016} remain valid under this more common implementation of ABC, but
they must be cast in terms of the limiting behaviour of the acceptance
probability $\alpha_{T}=\text{Pr}\left[  d\{\eta(\mathbf{y}),\eta
(\mathbf{z})\}\leq\varepsilon_{T}\right]  .$ Under this nearest-neighbour
interpretation, Corollary 1 in \citet{FMRR2016} demonstrates that consistency
requires $\alpha_{T}\rightarrow0$ as $T\rightarrow\infty$, and, in particular,
we require that $\alpha_{T}\asymp T^{-k_{\theta}/2}$, where $\asymp$ can be
understood as \textquotedblleft equal\textquotedblright\ in an order sense.
Moreover, for $N_{T}$ denoting the number of Monte Carlo draws used in ABC, it
must also be the case that $N_{T}\rightarrow\infty$ as $\alpha_{T}%
\rightarrow0$. To jointly satisfy these conditions we choose $N_{T}%
=500/\alpha_{T}$ and $\alpha_{T}=50T^{-3/2}$. In contrast, the number of MCMC
draws used to produce the exact predictives for each sample size remains fixed
at 20,000 draws, with a burn-in of 5,000\textbf{\ }iterations. However,
desp{ite the vast increase in the total number of Monte Carlo draws used in
ABC, as $T$ increases, the computation gains in using the ABC algorithm to
produce predictive distributions {remains marked}. {In accordance with the
result }}reported in {Section \ref{ma}}{{, for }$T=500$ the ABF {computation
is}} approximately {115 times faster {than the exact computation}. The
relative computational gain factors for $T=2000$ and $T=4000$ are 21 and 9,
respectively, while a gain o}f a factor of almost 5 is{\ still achieved at
$T=5000$.}\footnote{The requirement that $N_{T}$ diverge, at a particular
rate, is intimately related to the inefficient nature of the basic
accept/reject ABC approach. In large samples, it is often useful to use more
refined sampling techniques within ABC, as these approaches can often lead to
faster estimates of the ABC posterior than those obtained via the
accept/reject approach. Thus, at least in large samples, utilizing more
efficient ABC approaches will lead to a decrease in ABF computing times, which
will lead to an even higher computational gain over MCMC-based approaches. See
\cite{LF2016b} for alternative sampling schemes that only require
$N_{T}\rightarrow\infty$ very slowly.}

Before concluding we note that even though the predictive based on the
ABC\ posterior $p_{\varepsilon}(\theta|\eta^{(1)}(\mathbf{y}))$ does not
exhibit evidence of merging, as is clear from Panel D in Figure 2, for $T=500$
this approximate predictive is still very accurate as an estimate of the exact
predictive. Therefore, we conjecture that, in relatively small samples
Bayesian consistency may not be a necessary condition for ABC to yield
predictives that are close to the exact. However, the numerical merging
results demonstrate that this accuracy would degrade as the sample size
increased if the ABC\ posterior were not consistent.

\section{ABF in State Space Models\label{sec:latent}}

{So} far the focus has been on the case in which the vector of unknowns,
$\mathbf{\theta}$, is a $k_{\theta}$-dimensional set of parameters for which
informative summary statistics are sought for the purpose of {generating
probabilistic predictions.} By implication, and certainly in the case of {both
the INAR(1) and} MA(2) {examples}, the elements of $\mathbf{\theta}$ are
static in nature, with $k_{\theta}$ small enough for a set of summaries of
manageable dimension to be defined with relative ease.

{State space} models, in which the set of unknowns is augmented by a vector of
random parameters{ that is of dimension} equal to or greater than the sample
size, present additional challenges for ABC (\citealp{CREEL2015};
\citealp{Martin2017}), in terms of producing an ABC posterior for the static
parameters, $\mathbf{\theta}$, that is a good match for the exact. However,
the results in the previous section highlight that accuracy at the posterior
level is not necessary for agreement between the approximate and exact
predictives. This suggests that we may be able to choose a crude, but
computationally convenient, method of generating summaries for $\mathbf{\theta
}$ in a {state space} model, and still yield predictions that are close to
those given by exact methods. The results below confirm this intuition, as
well as making it clear that {exact} posterior inference on the {full}{ vector
of }states (and the extra computational complexities that such a procedure
entails) is not required for this accuracy to be achieved.

We illustrate these points in the context of a very simple {state space}
model, namely a stochastic volatility model for a financial return, $y_{t}$,
in which the logarithm of the random variance, $V_{t}$, follows a simple
autoregressive model of order 1 (AR(1)):%
\begin{align}
y_{t}  &  =\sqrt{V_{t}}\varepsilon_{t};\qquad\varepsilon_{t}\sim
i.i.d.N(0,1)\label{meas}\\
\ln V_{t}  &  =\theta_{1}\ln V_{t-1}+\eta_{t};\qquad\eta_{t}\sim
i.i.d.N(0,\theta_{2}) \label{state}%
\end{align}
with $\theta=(\theta_{1},\theta_{2})^{\prime}.$ {Prior specifications
$\theta_{1} \sim U(0.5,0.99)$ and $\theta_{2} \sim U(0.05,0.5)$ are employed.}
To generate summary statistics for the purpose of defining $p_{\varepsilon
}(\mathbf{\theta}|\mathbf{\eta}(\mathbf{y}))$, we begin by adopting the
following auxiliary generalized autoregressive conditional heteroscedastic
model with Gaussian errors ({GARCH-N}):%
\begin{align}
y_{t}  &  =\sqrt{V_{t}}\varepsilon_{t};\qquad\varepsilon_{t}\sim
i.i.d.N(0,1)\label{aux_y}\\
V_{t}  &  =\beta_{1}+\beta_{2}V_{t-1}+\beta_{3}y_{t-1}^{2}. \label{aux_v}%
\end{align}
As a computationally efficient summary statistic vector for use in ABF we use
the score of the {GARCH-N} likelihood function, computed using the simulated
and observed data, with both evaluated at the (quasi-) maximum likelihood
estimator of $\mathbf{\beta}=(\beta_{1},\beta_{2},\beta_{3})^{\prime}$ (see,
for example, \citealp{drovandi2015}, and \citealp{Martin2017}).

The exact predictive, $p(y_{T+1}|\mathbf{y})$, requires integration with
respect to both the static and latent parameters, including the value of the
latent variance at time $T+1$, $V_{T+1}.$ Defining $p(V_{T+1},\mathbf{V}%
,\mathbf{\theta}|\mathbf{y)}$ as the joint posterior for this full set of
unknowns (with $\mathbf{V}=(V_{1},V_{2},...,V_{T})^{\prime}$), and recognizing
the Markovian structure in the (log) variance process, we can represent this
predictive as%
\begin{align}
&  p(y_{T+1}|\mathbf{y})\nonumber\\
&  =\int_{V_{T+1}}\int_{\mathbf{V}}\int_{\mathbf{\theta}}p(y_{T+1}%
|V_{T+1})p(V_{T+1},\mathbf{V},\mathbf{\theta}|\mathbf{y)}d\mathbf{\theta
}d\mathbf{V}dV_{T+1}\nonumber\\
&  =\int_{V_{T+1}}\int_{\mathbf{V}}\int_{\mathbf{\theta}}p(y_{T+1}%
|V_{T+1})p(V_{T+1}|V_{T},\mathbf{\theta,y})p(\mathbf{V}|\mathbf{\theta
,y})p(\mathbf{\theta}|\mathbf{y})d\mathbf{\theta}d\mathbf{V}dV_{T+1}.
\label{exact_predict_sv}%
\end{align}
A {hybrid }Gibbs-{MH} MCMC algorithm is applied to yield posterior draws of
$\mathbf{\theta}$ and $\mathbf{V}$. We apply the sparse matrix sampling
algorithm of \cite{chan2009} to sample $\mathbf{V}$, and a standard Gibbs
algorithm to sample from the conditional posterior of $\mathbf{\theta}$ given
the states. Conditional on the draws of $\mathbf{\theta}$ and $V_{T}$ (in
particular), draws of $V_{T+1}$ and $y_{T+1}$ are produced {directly from
}$p(V_{T+1}|V_{T},\mathbf{\theta,y})$ {and }$p(y_{T+1}|V_{T+1})$
{respectively}, and the draws of $y_{T+1}$ used to produce an estimate of
$p(y_{T+1}|\mathbf{y}).$

Replacing $p(\mathbf{\theta}|\mathbf{y})$ in (\ref{exact_predict_sv}) by
$p_{\varepsilon}(\mathbf{\theta|\eta(y)})$, the approximate predictive is then
defined as%
\begin{align}
&  g(y_{T+1}|\mathbf{y})\label{approx_predict_sv}\\
&  =\int_{V_{T+1}}\int_{\mathbf{V}}\int_{\mathbf{\theta}}p(y_{T+1}%
|V_{T+1})p(V_{T+1}|V_{T},\mathbf{\theta,y})p(\mathbf{V}|\mathbf{\theta
,y})p_{\varepsilon}(\mathbf{\theta|\eta(y)})d\mathbf{\theta}d\mathbf{V}%
dV_{T+1}.\nonumber
\end{align}
In this case, however, draws are produced from $p_{\varepsilon}(\mathbf{\theta
|\eta(y)})$ via {the nearest neighbour version of }Algorithm 1 {(with
$\alpha=0.01$ and $N=50,000$)}, separately from the treatment of $\mathbf{V}.$
That is, posterior draws of $\mathbf{V}$, including $V_{T}$, are not an
automatic output of a simulation algorithm applied to the joint set of
unknowns $\mathbf{\theta}$ and $\mathbf{V}$, as was the case in the estimation
of (\ref{exact_predict_sv}). However, the estimation of $g(y_{T+1}%
|\mathbf{y})$ requires only that posterior draws of $V_{T}$ and
$\mathbf{\theta}$ are produced; that is, posterior inference on the
\textit{full vector }$\mathbf{V}$, as would require a backward {sampling} step
to be embedded within the simulation algorithm, is not necessary. All that is
required is that $\mathbf{V}_{1:T-1}$ are integrated out, and this can occur
via a forward filtering step alone. The implication of this is that,
conditional on a simple $i.i.d$. version of Algorithm 1 being adopted (i.e.,
that no MCMC modifications of ABC are employed), a simulation-based estimate
of the approximate predictive can {still }be produced using $i.i.d$. draws
only. As such, the great gains in computational speed afforded by the use of
ABC - including access to parallelization - continue to obtain even when
latent variables characterize the true DGP.

\begin{figure}[ptb]
\setlength\figureheight{3.250cm}
\setlength\figurewidth{5.75cm}
\input{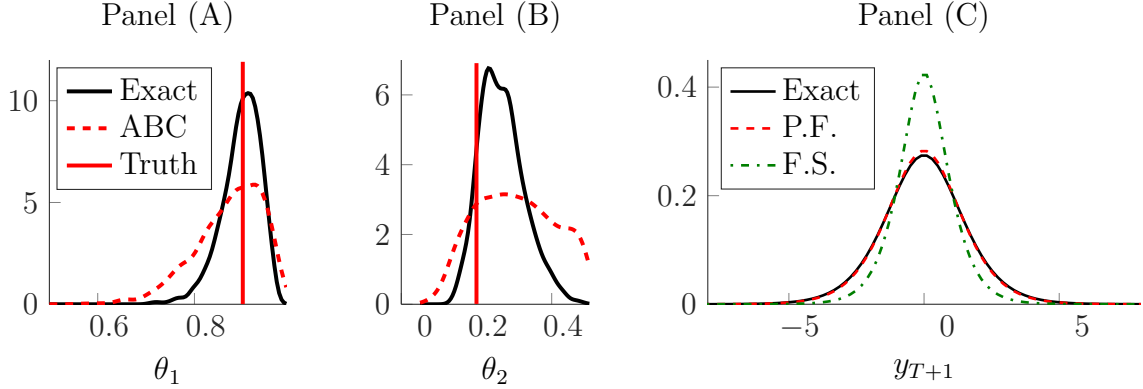}\caption{Panels (A) and (B) depict the marginal posteriors
(exact and ABC) for $\theta_{1}$ and $\theta_{2}$, respectively. Panel (C)
plots the one-step-ahead predictive density functions - both exact and
approximate (ABC-based). P.F. indicates the approximate predictive computed
using the particle filtering step; F.S. indicates the approximate predictive
computed using a forward simulation step (for the latent variance) only. The
red vertical line (denoted by `Truth' in the key) represents the true value of
the relevant parameter in Panels (A) and (B).}%
\label{postgarch}%
\end{figure}

Panels (A) and (B) of Figure \ref{postgarch} depict the marginal ABC
posteriors of $\theta_{1}$ and $\theta_{2}$\ alongside the MCMC-based
comparators. The {dashed} curve in Panel (C) of Figure \ref{postgarch} then
represents {the estimate} of (\ref{approx_predict_sv}), in which the particle
filter is used to integrate out the latent variances, and the{ full curve}
represents the MCMC-based estimate of (\ref{exact_predict_sv}). As is
consistent with the numerical results recorded {earlier }for the {INAR(1) and
}MA(2) {examples}, the difference between the\textbf{ }approximate and exact
posteriors is marked, whilst - at the same time - the approximate predictive
is almost equivalent to the exact, and having been produced using a much
simpler algorithm, and in a fraction of the time.

{The importance of the particle filtering step} in obtaining this (near)
equivalence is highlighted by the inclusion of a third predictive (the
{dot-dashed curve}) in Panel (C) of Figure \ref{postgarch}, which is
constructed by replacing the particle filtering step by a sim{ple forward
simulation of the latent variance model in (\ref{state}) - conditional on the
ABC draws of }$\mathbf{\theta}$ -\textbf{\ }such that inference on $V_{T}$ is
itself conditioned on $\mathbf{\eta}(\mathbf{y})$, rather than $\mathbf{y}.$
Without full posterior inference on $V_{T}$, the gains obtained by ABC
inference on $\mathbf{\theta}$ (in terms of computational speed and ease) are
achieved only at the cost of producing an inaccurate estimate of the exact
predictive. We reiterate, however, that full posterior inference on $V_{T}$
({as reflected in the very accurate dashed} curve in Panel (C) of {Figure
\ref{postgarch}}) requires \textit{only }a particle filtering step.

\begin{figure}[ptb]
\centering
\setlength\figureheight{3.250cm}
\setlength\figurewidth{14.5cm}
\input{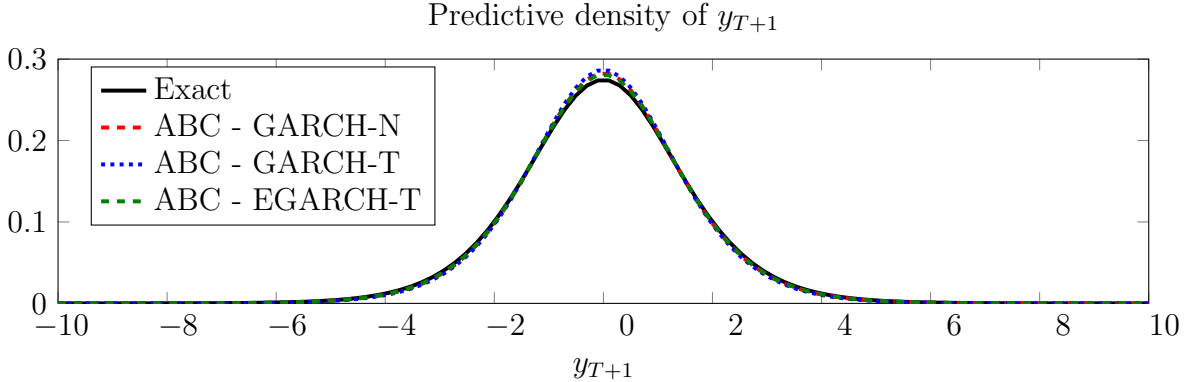}\caption{Exact and approximate (ABC-based) predictives.
{The three approximate posteriors are based on the auxiliary models indicated
in the key}. All approximate predictives use the particle filtering step. }%
\label{predaux}%
\end{figure}

To further highlight the apparent second-order importance of \textit{static}
parameter inference on the predictive, along with the exact and approximate
predictives reproduced in Panel (C) of Figure \ref{postgarch} ({namely the
full and dashed curves}), Figure \ref{predaux} plots two alternative
approximate predictives that use different auxiliary models to define the
summary statistics. The {GARCH-T} auxiliary model employs the structure as
defined in (\ref{aux_y}) and (\ref{aux_v}), but with a Student-t error term,
$\varepsilon_{t}\sim i.i.d.\;t(\nu)$, used to accommodate extra leptokurtosis
in the return. The {EGARCH-T} auxiliary model also employs Student-t errors,
but with skewness in the {return modeled} via an asymmetric specification for
the conditional variance:
\[
\ln V_{t}=\beta_{0}+\beta_{1}\ln V_{t-1}+\beta_{2}\left(  \left\vert
\varepsilon_{t-1}\right\vert -\mathbb{E}(\left\vert \varepsilon_{t-1}%
\right\vert )\right)  +\beta_{3}\varepsilon_{t-1}.
\]
As is clear, given the inclusion of the particle filtering step, the choice of
auxiliary model ({and hence summary statistics) underpinning }$p_{\varepsilon
}(\mathbf{\theta|\eta(y)})$ has little impact on the nature of the resultant
predictive, with all three auxiliary models generating approximate predictives
that are extremely close to the exact.

This robustness of prediction to the choice of summary statistics augers well
for the automated use of ABC as a method for generating Bayesian predictions
in models where {finely}-tuned, specialized algorithms have been viewed as an
essential ingredient up to now. It also suggests that Bayesian predictions
that are close to exact can be produced in models in which exact prediction is
infeasible, {that is}, in models where the DGP - and hence, the exact
predictive itself - is unavailable. It is precisely such a case that we
explore in the following empirical section, with performance now gauged not in
terms of the accuracy with which any particular $g(y_{T+1}|\mathbf{y})$
matches $p(y_{T+1}|\mathbf{y})$, but in terms of out-of-sample predictive accuracy.

\section{Empirical Illustration: Forecasting Financial Returns and Volatility
\label{numerical}}

\subsection{Background, model and computational details}

The effective management of financial risk entails the ability to plan for
unexpected, and potentially large, movements in asset prices. Central to this
is the ability to accurately quantify the probability distribution of the
future return on the asset, including its degree of variation, or volatility.
The stylized features of time-varying and autocorrelated volatility, allied
with non-Gaussian return distributions, are now extensively documented in the
literature ({\citealp{bollerslev1992arch}); with more recent work focusing
also on random `jump' processes, both in the asset price itself and its
volatility (\citealp{broadie2007model}; \citealp{bandi2016};
\citealp{maneesoonthorn2017}). Empirical regularities documented in the option
pricing literature (see \citealp{garcia2011estimation}} {for a review}), most
notably implied volatility `smiles', are also viewed as evidence that asset
prices do not adhere to the geometric Brownian motion assumption underlying
the ubiquitous Black-Scholes option price, and that the processes driving
asset returns are much more complex in practice.

Motivated by these now well-established empirical findings, we explore here a
state space specification for financial returns on the S\&P500 index, in which
both stochastic volatility \textit{and} random jumps are accommodated. To do
so, we supplement a measurement equation for the daily return, in which a
dynamic jump process features, with a second measurement equation based on
bipower variation, {constructed using} five-minute intraday returns over the
trading day (\citealp{barndorff2004power}). Such a model is representative of
models used recently to capture returns data in which \textit{clustering} of
jumps features, in addition to the stylized autocorrelation in the diffusive
variance (\citealp{fulop2014self}; \citealp{ait2015modeling};
\citealp{bandi2016}; \citealp{maneesoonthorn2017}). It also reflects the
recent trend of exploiting high frequency data to construct - and use as
additional measures in state space settings - nonparametric measures of return
variation, including jumps therein (\citealp{koopman2012analysis};
\citealp{maneesoonthorn2012probabilistic}; \citealp{maneesoonthorn2017}). To
capture the possibility of extreme movements in volatility, and in the spirit
of \cite{lombardi2009indirect} and \cite{Martin2017}, we adopt an $\alpha
${-stable process for the }volatility innovations.{ Despite the lack of a
closed-form transition density, the }$\alpha$-stable process presents no
challenges for ABC-based inference and forecasting, given that such a process
can still be simulated via the algorithm of \cite{chambers1976method}.

In summary, the assumed data generating process comprises two measurement
equations: one {based on }daily logarithmic returns, $r_{t}$,
\begin{equation}
r_{t}=\exp\left(  \frac{h_{t}}{2}\right)  \varepsilon_{t}+\Delta N_{t}Z_{t},
\label{ret}%
\end{equation}
where $\varepsilon_{t}\sim i.i.d.N(0,1)$, $h_{t}$\ denotes the latent
logarithmic variance process, $\Delta N_{t}$\ the latent jump occurrence and
$Z_{t}$\ the latent jump size; {and a second using} logarithmic bipower
variation,
\begin{equation}
\ln BV_{t}=\psi_{0}+\psi_{1}h_{t}+\sigma_{BV}\zeta_{t}, \label{bv}%
\end{equation}
{where }$BV_{t}=\frac{\pi}{2}\left(  \frac{M}{M-1}\right)  \sum_{i=2}%
^{M}\left\vert r_{t_{i}}\right\vert \left\vert r_{t_{i-1}}\right\vert $, with
$r_{t_{i}}$ denoting the $i^{th}$, of $M$ equally-spaced returns observed
during day $t$, {and} $\zeta_{t}\sim i.i.d.N(0,1)$. As is now well-known
(\citealp{barndorff2004power}), under certain conditions $BV_{t}$ {is a
consistent, but potentially biased ({for finite }}$M$){, estimate of
integrated volatility over day }$t$, with $h_{t}$ here {being a discretized
representation of the (logarithm of the) latter.} The latent states in
equations \eqref{ret} and \eqref{bv}, $h_{t}$, $Z_{t}$ and $\Delta N_{t}$,
evolve, respectively, according to{
\begin{align}
h_{t}  &  =\omega+\rho h_{t-1}+\sigma_{h}\eta_{t}\label{ht}\\
Z_{t}  &  \sim N(\mu,\sigma_{z}^{2})\label{zt}\\
\text{Pr}(\Delta N_{t}=1|\mathcal{F}_{t-1})  &  =\delta_{t}=d+\beta
\delta_{t-1}+\gamma\Delta N_{t-1}, \label{delnt}%
\end{align}
where }$\eta_{t}\sim i.i.d.\mathcal{S}(\alpha,-1,0,dt=1)$. {We note that the
model for the jump intensity, }$\delta_{t}$, is the conditionally
deterministic Hawkes structure adopted by \cite{fulop2014self},
\cite{ait2015modeling} and \cite{maneesoonthorn2017}. We {estimate} $d$ (in
\eqref{delnt}) {indirectly }via the unconditional intensity implied by this
particular structure, namely, $d^{0}=d/(1-\beta-\gamma).$

Exact inference on the full set of static parameters,
\begin{equation}
\mathbf{\theta}=(\psi_{0},\psi_{1},\sigma_{BV},\omega,\rho,\sigma_{h}%
,\alpha,d^{0},\beta,\gamma,\mu,\sigma_{z})^{\prime}, \label{eq:theta}%
\end{equation}
is challenging, not only due to the overall complexity of the model, but in
particular as a consequence of the presence of $\alpha$-stable (log)
volatility transitions. Hence, ABC is a natural choice for inference on
$\mathbf{\theta}$. Moreover, given the previously presented evidence regarding
the accuracy with which ABC-based predictives match the predictive that
\textit{would} be yielded by an exact method, one proceeds with some
confidence to build Bayesian predictives via ABC posteriors.

To measure the predictive performance of our ABF approach, we consider an
out-of-sample predictive exercise, whereby we assess the {relative }accuracy
of {approximate predictives based on alternative} choices of summaries,
$\eta(\mathbf{y})$.\footnote{In contrast to the previous numerical examples,
where $y_{t}$ was univariate, in this example our goal is to jointly forecast
log-returns, $r_{t}$, and the logarithm of bi-power variation, $\ln BV_{t}$.
Therefore, in what follows $y_{T+1}=(r_{T+1},\ln BV_{T+1})^{\prime}$.} We make
two comments here. First, and as highlighted in the previous section, a
forward particle filtering step (conditional on draws from the ABC posterior)
is required to produce the full posterior inference on the latent state,
$h_{T},$ {that is}, {in turn}, required to construct $g(y_{T+1}|\mathbf{y})$
under any choice for $\eta(\mathbf{y}).$ We adopt the bootstrap particle
filter of \cite{gordon1993novel} for this purpose.\footnote{Note that the
conditionally deterministic structure in (\ref{delnt}) means that no
additional filtering step is required in order to model the jump intensity at
time $T.$} Second, when the {data generating process} is correctly specified,
and if the conditions for Bayesian consistency and asymptotic normality of
both the exact and ABC posteriors are satisfied, then the out-of-sample
accuracy of $g(y_{T+1}|\mathbf{y})$ is bounded above by that of $p(y_{T+1}%
|\mathbf{y})$, as measured by some proper scoring rule, as shown in Section
3.3. Hence, in {choosing} $\eta(\mathbf{y})$, {from a set of alternatives,
according to the accuracy of the associated predictive,} we are {- in spirit -
choosing} an approximate predictive that is as close as possible (in terms of
predictive accuracy) to the inaccessible exact predictive.

We consider observed data from 26 February 2010 to 7 February 2017, comprising
1750 {daily }observations {on both }$r_{t}$ and $BV_{t}$. We reserve the most
recent 250 observations ({approximately one trading year}) for one-step-ahead
predictive assessments, {using an expanding window approach. }In the spirit of
{the }{preceding section}, we {implement ABC using} the scores of {alternative
}auxiliary GARCH models {fitted to daily returns}. In this case, however, we
must also conduct inference on the {parameters of the additional measurement
equation, (\ref{bv}),} and the jump processes in (\ref{zt}) and (\ref{delnt});
hence we supplement the auxiliary model scores with additional summary
statistics based on {both }$BV_{t}$ as well as the realized jump variation
measure, $JV_{t}=\max(RV_{t}-BV_{t},0)$, where $RV_{t}=\sum_{i=1}^{M}r_{t_{i}%
}$ defines so-called realized variance {for day }$t$.

We consider four auxiliary models: GARCH with normal and\ Student-t errors
(GARCH-N and GARCH-T, respectively), threshold GARCH with Student-t errors
(TARCH-T), and the realized GARCH (RGARCH) model of \cite{hansen2012realized}.
Table \ref{tab:auxSpecs} {details} {these four} models, {plus} the {additional
summary statistics that we employ} {in each case}. {In particular, we note}{
that the RGARCH model itself incorporates a component in which }$\ln BV_{t}$
{is modeled; hence, {in this case} we do not adopt additional summary
statistics based on this measure. }We {adopt} {independent }uniform priors
{for} all static parameters {in the structural model}, {subject to relevant
model-based restrictions, }with the lower and upper bounds for each given in
Table \ref{tab:prior}. {All ABC posteriors are produced by the nearest
neighbour version of Algorithm \ref{ABC} described in Section \ref{inar_num},
but with $\alpha_{T}$ and $N_{T}$ depending on the sample size $T$ as per
\citet{FMRR2016} {(see Section
\ref{numerics_consistency} for additional discussion.)}} {We note that the use
}of GARCH auxiliary models ability to yield summary statistics that guarantee
posterior concentration has been numerically verified in {similar models} in
\cite{Martin2017}. However, we believe a formal verification {of Bayesian
consistency in the current context}, as was {done} with the examples in
Section 3.4, is beyond the scope of this paper.

\subsection{Empirical forecasting results}

In Table \ref{tab:post} we report the marginal ABC posterior means (MPM) and
the 95\% highest posterior density (HPD) intervals for the elements of
$\theta$, based on the four choices of {summaries}. The posterior results
obtained {via the} {first three sets (based, in turn, on the }GARCH-N, GARCH-T
and TARCH-T auxiliary models) are {broadly }similar, except for the TARCH-T
auxiliary model {producing noticeable narrower }95\% HPD intervals {for}
$\omega$, $\mu$ and $\sigma_{z}$ than the other auxiliary models. In contrast
to the relative conformity of these three sets of results, the RGARCH
auxiliary model ({augmented by the} {additional} summaries) produces ABC
posteriors that differ quite substantially. {Most notably}, and with reference
to the latent process for $h_{t}$ in (\ref{ht}), ABC based on {this fourth set
of summaries} produces a{ larger }MPM for $\omega$, a lower MPM of $\rho$, and
a {smaller }MPM {for} $\sigma_{h}$ {than do the other instances of ABC}. {In
addition, this version produces a larger point estimate for the }mean jump
size, $\mu$, {plus a smaller point estimate of} the jump size variation,
$\sigma_{z}$. These differences imply somewhat different conclusions regarding
the process generating returns than those implied by the other three sets of
ABC posterior results. As a consequence there would be differing degrees of
concordance between the four sets of ABC posteriors and the corresponding
exact, unattainable, posteriors. The question of interest here is the extent
to which such differences translate into substantial differences at the
predictive level, where a judgment is made solely in terms of out-of-sample
predictive accuracy, given our {lack of access} to the exact predictive.

\begin{table}[ptb]
\caption{Auxiliary model specifications for ABC posterior inference for the
model in (\ref{ret})-(\ref{delnt}). The error terms, $\varepsilon_{t}$ and
$\zeta_{t}$, {in the second and third columns }are specified as \textit{i.i.d}%
. The notation $\hat{\sigma}_{t}$ in the third column refers to fitted
volatility from the corresponding volatility equation in the auxiliary model.
The final column gives the set of supplementary summary statistics used {in
addition to} the scores from each auxiliary model. {The total number of
summary statistics }used in each specification is denoted by $d_{\eta}$ in the
first column.\bigskip}%
\label{tab:auxSpecs}%
\centering%
\begin{tabular}
[c]{lll}\hline\hline
\multicolumn{2}{c}{Auxiliary Model} & Supplementary Statistics\\\hline\hline
GARCH-N & $r_{t}=\sigma_{t}\varepsilon_{t}$, $\varepsilon_{t}\sim N(0,1)$ &
$Mean(sign(r_{t})\sqrt{JV_{t}})$, $Var(JV_{t})$\\
$d_{\eta}=11$ & $\sigma_{t}^{2}=\gamma_{0}+\gamma_{1}r_{t-1}^{2}+\gamma
_{2}\sigma_{t-1}^{2}$ & $Corr(JV_{t},JV_{t-1})$\\
&  & Skewness($\ln BV_{t}$), Kurtosis($\ln BV_{t}$)\\
&  & Regression coefficients from\\
&  & $\ln BV_{t}=\kappa_{0}+\kappa_{1}\ln\hat{\sigma}_{t}^{2}+\kappa_{3}%
\zeta_{t}$\\\hline
GARCH-T & $r_{t}=\sigma_{t}\varepsilon_{t}$, $\varepsilon_{t}\sim t(\nu)$ &
$Mean(sign(r_{t})\sqrt{JV_{t}})$, $Var(JV_{t})$\\
$d_{\eta}=12$ & $\sigma_{t}^{2}=\gamma_{0}+\gamma_{1}r_{t-1}^{2}+\gamma
_{2}\sigma_{t-1}^{2}$ & $Corr(JV_{t},JV_{t-1})$\\
&  & Skewness($\ln BV_{t}$), Kurtosis($\ln BV_{t}$)\\
&  & Estimated regression coefficients from:\\
&  & $\ln BV_{t}=\kappa_{0}+\kappa_{1}\ln\hat{\sigma}_{t}^{2}+\kappa_{3}%
\zeta_{t}$\\\hline
TARCH-T & $r_{t}=\sigma_{t}\varepsilon_{t}$, $\varepsilon_{t}\sim t(\nu)$ &
$Mean(sign(r_{t})\sqrt{JV_{t}})$, $Var(JV_{t})$\\
$d_{\eta}=13$ & $\sigma_{t}^{2}=\gamma_{0}+\gamma_{1}r_{t-1}^{2}+\gamma
_{2}I_{(r_{t-1}<0)}r_{t-1}^{2}$ & $Corr(JV_{t},JV_{t-1})$\\
& $+\gamma_{3}\sigma_{t-1}^{2}$ & Skewness($\ln BV_{t}$), Kurtosis($\ln
BV_{t}$)\\
&  & Estimated regression coefficients from:\\
&  & $\ln BV_{t}=\kappa_{0}+\kappa_{1}\ln\hat{\sigma}_{t}^{2}+\kappa_{3}%
\zeta_{t}$\\\hline
RGARCH & $r_{t}=\sigma_{t}\varepsilon_{t}$, $\varepsilon_{t}\sim N(0,1)$ &
$Mean(sign(y_{t})\sqrt{JV_{t}})$, $Var(JV_{t})$\\
$d_{\eta}=12$ & $\ln\sigma_{t}^{2}=\gamma_{0}+\gamma_{1}\ln BV_{t-1}%
+\gamma_{2}\ln\sigma_{t-1}^{2}$ & $Corr(JV_{t},JV_{t-1})$, Kurtosis($\ln
BV_{t}$)\\
& $\ln BV_{t}=\gamma_{3}+\gamma_{4}\ln\sigma_{t-1}^{2}+\gamma_{5}%
\varepsilon_{t}$ & \\
& $+\gamma_{6}\left(  \varepsilon_{t}^{2}-1\right)  +\gamma_{7}u_{t}$,
$u_{t}\sim N(0,1)$ & \\\hline
\end{tabular}
\end{table}

\begin{table}[ptb]
\caption{Lower and upper bounds on the uniform prior specifications used for
each element of $\theta$, as defined in (\ref{eq:theta}).\bigskip}%
\label{tab:prior}%
\begin{tabular}
[c]{rrrrrrrrrrrrr}\hline\hline
Parameter & $\psi_{0}$ & $\psi_{1}$ & $\sigma_{BV}$ & $\omega$ & $\rho$ &
$\sigma_{h}$ & $\alpha$ & $d$ & $\beta$ & $\gamma$ & $\mu$ & $\sigma_{z}%
$\\\hline\hline
Lower & -0.50 & 0.50 & 0.001 & -1 & 0.50 & 0.001 & 1.50 & 0.001 & 0.50 &
0.001 & -1 & .50\\
Upper & 0.50 & 1.50 & 1 & 1 & 0.99 & 0.30 & 2 & 0.30 & 0.99 & 0.20 & 1 &
3\\\hline
\end{tabular}
\end{table}

\begin{table}[ptb]
\caption{Marginal posterior means (MPM) and 95\% highest posterior density
(HPD) intervals for each of the elements of $\theta$, as defined in
(\ref{eq:theta}), obtained from ABC posterior inference using the four
auxiliary models and supplementary statistics defined in Table
\ref{tab:auxSpecs}.\bigskip}%
\label{tab:post}%
\centering%
\begin{tabular}
[c]{lllllllll}\hline\hline
& \multicolumn{2}{c}{GARCH-N} & \multicolumn{2}{c}{GARCH-T} &
\multicolumn{2}{c}{TARCH-T} & \multicolumn{2}{c}{RGARCH}\\
& MPM & 95\% HPD & MPM & 95\% HPD & MPM & 95\% HPD & MPM & 95\%
HPD\\\hline\hline
$\psi_{0}$ & \multicolumn{1}{r}{-0.02} & {\small (-0.47,0.47)} &
\multicolumn{1}{r}{0.00} & {\small (-0.49,0.46)} & \multicolumn{1}{r}{-0.01} &
{\small (-0.47,0.48)} & \multicolumn{1}{r}{-0.01} & {\small (-0.48,0.47)}\\
$\psi_{1}$ & \multicolumn{1}{r}{1.26} & {\small (0.77,1.49)} &
\multicolumn{1}{r}{1.25} & {\small (0.83,1.49)} & \multicolumn{1}{r}{1.20} &
{\small (0.73,1.49)} & \multicolumn{1}{r}{0.96} & {\small (0.51,1.45)}\\
$\sigma_{BV}$ & \multicolumn{1}{r}{0.45} & {\small (0.02,0.96)} &
\multicolumn{1}{r}{0.47} & {\small (0.03,0.95)} & \multicolumn{1}{r}{0.48} &
{\small (0.02,0.95)} & \multicolumn{1}{r}{0.55} & {\small (0.04,0.99)}\\
$\omega$ & \multicolumn{1}{r}{-0.04} & {\small (-0.68,0.38)} &
\multicolumn{1}{r}{-0.10} & {\small (-0.34,0.20)} & \multicolumn{1}{r}{-0.17}
& {\small (-0.48,-0.01)} & \multicolumn{1}{r}{0.19} & {\small (-0.95,0.97)}\\
$\rho$ & \multicolumn{1}{r}{0.94} & {\small (0.81,0.99)} &
\multicolumn{1}{r}{0.93} & {\small (0.83,0.99)} & \multicolumn{1}{r}{0.92} &
{\small (0.81,0.98)} & \multicolumn{1}{r}{0.79} & {\small (0.52,0.99)}\\
$\sigma_{h}$ & \multicolumn{1}{r}{0.20} & {\small (0.08,0.29)} &
\multicolumn{1}{r}{0.21} & {\small (0.08,0.30)} & \multicolumn{1}{r}{0.20} &
{\small (0.06,0.30)} & \multicolumn{1}{r}{0.13} & {\small (0.01,0.29)}\\
$\alpha$ & \multicolumn{1}{r}{1.76} & {\small (1.52,1.98)} &
\multicolumn{1}{r}{1.76} & {\small (1.52,1.99)} & \multicolumn{1}{r}{1.77} &
{\small (1.52,1.99)} & \multicolumn{1}{r}{1.80} & {\small (1.52,1.99)}\\
$d^{0}$ & \multicolumn{1}{r}{0.11} & {\small (0.01,0.28)} &
\multicolumn{1}{r}{0.11} & {\small (0.01,0.27)} & \multicolumn{1}{r}{0.10} &
{\small (0.01,0.28)} & \multicolumn{1}{r}{0.10} & {\small (0.01,0.27)}\\
$\beta$ & \multicolumn{1}{r}{0.69} & {\small (0.51,0.90)} &
\multicolumn{1}{r}{0.69} & {\small (0.51,0.91)} & \multicolumn{1}{r}{0.69} &
{\small (0.51,0.90)} & \multicolumn{1}{r}{0.69} & {\small (0.52,0.90)}\\
$\gamma$ & \multicolumn{1}{r}{0.12} & {\small (0.02,0.20)} &
\multicolumn{1}{r}{0.12} & {\small (0.02,0.20)} & \multicolumn{1}{r}{0.12} &
{\small (0.02,0.20)} & \multicolumn{1}{r}{0.13} & {\small (0.03,0.20)}\\
$\mu$ & \multicolumn{1}{r}{0.07} & {\small (-0.87,0.94)} &
\multicolumn{1}{r}{0.05} & {\small (-0.86,0.90)} & \multicolumn{1}{r}{0.12} &
{\small (-0.81,0.88)} & \multicolumn{1}{r}{0.23} & {\small (-0.69,0.94)}\\
$\sigma_{z}$ & \multicolumn{1}{r}{1.21} & {\small (0.52,2.57)} &
\multicolumn{1}{r}{1.23} & {\small (0.53,2.72)} & \multicolumn{1}{r}{1.14} &
{\small (0.53,2.49)} & \multicolumn{1}{r}{1.01} & {\small (0.53,2.15)}\\\hline
\end{tabular}
\end{table}

To summarize predictive performance over the out-of-sample period, {average
LS, QS and CRPS values for each} of the four approximate predictives are
reported in Table \ref{tab:outscores}, with the {largest figure in each case
indicated in bold}. The results {indicate} that the predictive distribution
{for }$r_{t}$ generated {via} the TARCH-T auxiliary model ({and additional
summaries) }performs best {according to} all three score criteria. {The}
GARCH-N auxiliary model ({and additional summaries) }generates the
{best-performing} predictive distribution for $\ln BV_{t}$ according to{\ LS
and QS}, {but }with CRPS still suggesting that the TARCH-T-based predictive
performs the best. It is interesting to note that {the set of statistics that
generates }the worst overall predictive {performance } (with the lowest
predictive scores in all but one case) is {that which includes }the RGARCH
{auxiliary }model - i.e. the {set} that resulted in {ABC }marginal posteriors
that were distinctly different from those obtained via the {other three
statistic sets}.

\begin{table}[ptb]
\caption{Average predictive log score (LS), quadratic score (QS) and
cumulative rank probability score (CRPS) for the one-step-ahead {approximate
}predictive distributions of $r_{t}$ and $\ln BV_{t}$, evaluated between 11
February 2016 and 7 February 2017. The figures in bold indicate the largest
{average }score amongst the four sets of summaries.\bigskip}%
\label{tab:outscores}%
\centering
\begin{tabular}
[c]{rlrrrr}\hline\hline
&  & \multicolumn{1}{l}{GARCH-N} & \multicolumn{1}{l}{GARCH-T} &
\multicolumn{1}{l}{TARCH-T} & \multicolumn{1}{l}{RGARCH}\\\hline\hline
& LS & -1.57 & -1.28 & \textbf{-1.20} & -1.95\\
\multicolumn{1}{l}{$r_{t}$} & QS & 0.38 & 0.47 & \textbf{0.52} & 0.27\\
& CRPS & -1.52 & -1.05 & \textbf{-0.99} & -2.10\\\hline
& LS & \textbf{-2.73} & -2.76 & -2.93 & -2.83\\
\multicolumn{1}{l}{$\ln BV_{t}$} & QS & \textbf{0.10} & 0.05 & 0.02 & 0.09\\
& CRPS & -2.04 & -1.42 & \textbf{-1.38} & -2.57\\\hline
\end{tabular}
\end{table}

{In summary, these predictive outcomes - in which the{ approximate predictive
produced using the}\textit{ }T-GARCH-based set of summaries performs best -
suggest that this choice of summaries be the one settled upon. {Repeating the
point made above, for any finite sample the predictive performance of any
approximate predictive will ({under appropriate regularity conditions}}}) {{be
bounded above by that of the exact predictive; however, this difference is
likely to be minor under correct specification of the DGP.} }

\section{Discussion}

{This paper explores} the use of approximate Bayesian computation (ABC) in
generating probabilistic forecasts and{ {proposes }the concept of} approximate
Bayesian forecasting (ABF). Theoretical and numerical evidence has been
presented which indicates that if the assumed {data generating process (DGP)}
is correctly specified, very little is lost - in terms of forecast accuracy -
by conducting approximate inference (only) on the unknowns that characterize
the DGP. A caveat here applies to latent variable models, in that exact
inference on the {conditioning latent }state(s) would appear to be important.
However, even that requires only independent particle draws, to supplement the
computationally fast and simple independent draws of the static parameters via
ABC; detracting little from the overall conclusion that ABC represents a
powerful base on which to produce accurate Bayesian forecasts in a short
amount of time. Whilst the asymptotic results based on merging formally
exploit the property of Bayesian consistency, numerical evidence suggests that
lack of consistency for the ABC posteriors does not preclude the possibility
of a close match to the exact predictive being produced {in any given finite
sample}.{ The theoretical results presented regarding expected scores }are{
also borne out in the numerical illustrations, with minor - if any -
forecasting loss incurred by moving from exact to approximate prediction, for
the sample sizes considered.}

Importantly, in an empirical setting where the exact predictive is
unattainable, the idea of choosing ABC summaries to produce the best
performing approximate predictive is a sensible approach to adopt when
predictive accuracy is the primary goal, and when the true {DGP} is of course
unknown. What remains the subject of on-going investigation by the authors, is
the interplay between new results on the impact on ABC inference of model
misspecification (\citealp{frazier2017model}) and the performance of ABC in a
forecasting setting in which misspecification {of the DGP} is explicitly
acknowledged. The outcomes of this exploration are reserved for future
research output.

\bibliographystyle{apalike}
\bibliography{refs_fcast}

\appendix

\section{Proofs}

\label{thm2_proof} Let $\{\mathcal{F}_{t}:t\geq0\}$ be a filtration associated
with the probability space $(\Omega,\mathcal{F},\mathbb{P})$. {The sequence
$\{y_{t}\}_{t\geq1}$ is adapted to the filtration $\{\mathcal{F}_{t}\}$. Let
$P(\cdot|{\mathbf{\theta}})$ denote the generative model for $\mathbf{y}$.
Define
\[
F_{\mathbf{y}}=P(\cdot|\mathbf{\theta}_{0},\mathbf{y})
\]
to be the true conditional predictive distribution.}

Throughout the remainder, let $y_{T+1}$ denote a point of support for the
random variable $Y_{T+1}$. Recall the definitions \begin{flalign*}
P_{\mathbf{y}}&=\int_{\Theta} P(\cdot|\mathbf{\theta
},\mathbf{y})d\Pi[\mathbf{\theta}|\mathbf{y}],\;\;
G_{\y}=  \int_{\Theta} P(\cdot|\mathbf{\theta
},\mathbf{y})d\Pi[\mathbf{\theta}|\mathbf{\eta}(\mathbf{y})].%
\end{flalign*}

The results of this section hold under the following high-level assumptions.
Lower level sufficient conditions for these assumptions can easily be given,
however, such a goal is not germane to the discussion at hand.

\begin{assumption}
\label{ass1} The following are satisfied: (1) $p(\mathbf{y}|\mathbf{\theta})$
is $\mathcal{F}_{T}$ measurable for all $\mathbf{\theta}\in\Theta$ and for all
$T\geq1$; (2) For all $\mathbf{\theta}\in\Theta$ and all $T\geq1$,
$0<p(\mathbf{y}|\mathbf{\theta})<\infty$; (3) There exists a unique
$\theta_{0}\in\Theta$, such that $\mathbf{y}\sim P(\cdot|{\mathbf{\theta}_{0}%
})\in\mathbb{P}$; (4) For any $\epsilon>0$, and $A_{\epsilon}%
:=\{\mathbf{\theta}\in\Theta:\Vert\mathbf{\theta}-\mathbf{\theta}_{0}%
\Vert>\epsilon\}$, $\Pi\lbrack A_{\epsilon}|\mathbf{y}]\rightarrow
_{\mathbb{P}}0$ and $\Pi\lbrack A_{\epsilon}|\mathbf{\eta}(\mathbf{y}%
)]\rightarrow_{\mathbb{P}}0$, i.e. Bayesian consistency of $\Pi\lbrack
A_{\epsilon}|\mathbf{y}]$ and $\Pi\lbrack A_{\epsilon}|\eta(\mathbf{y})]$ holds.
\end{assumption}

\subsection{Lemma}

We begin the proof by first showing the following Lemma.

\noindent\textbf{Lemma} \textit{$G_{\mathbf{y}}$ is a conditional measure and
$G_{\mathbf{y}}(\Omega)=1$.}

\begin{proof}
The result follows by verifying the required conditions for a probability measure.

\noindent(1) For any $B\in\mathcal{F}$, $1\!\mathrm{l}[Y\in B]g(Y|\mathbf{y}%
)\geq0$ and hence
\[
G_{\mathbf{y}}(B)=\int_{\Omega}1\!\mathrm{l}[Y\in B]g(Y|\mathbf{y})dY\geq0.
\]

\noindent(2) By definition, $G_{\mathbf{y}}(\{\emptyset\})=0.$

\noindent(3) Let $E_{k}=[a_{k},b_{k})$, $k\geq1$, be a collection of disjoint
sets (in $\mathcal{F}$). By construction, for all $\omega\in\Omega$,
$1\!\mathrm{l}[Y\in E_{k}]g(Y|\mathbf{y}(\omega))\geq0$ and hence
\begin{flalign*}
G_{\y}\left(\bigcup_{k=1}^{\infty} E_{k}\right)&=\int \1\left[Y\in \bigcup_{k=1}^{\infty} E_{k}\right]g(Y|\mathbf{y})dY=\int\sum_{k=1}^{\infty}\1[Y\in E_{k}]g(Y|\mathbf{y})dY\\&=\sum_{k=1}^{\infty}\int\1[Y\in E_{k}]g(Y|\mathbf{y})dY,
\end{flalign*}where the last line follows by Fubini's theorem.

\noindent(4) All that remains to be shown is that $G_{\mathbf{y}}(\Omega)=1$.
By definition
\[
G_{\mathbf{y}}(\Omega)=\int_{\Omega}g(Y|\mathbf{y})dY=\int_{\Omega}%
\int_{\Theta}p(Y|\theta,\mathbf{y})d\Pi\lbrack\mathbf{\theta}|\eta
(\mathbf{y})]dY.
\]
By Fubini's Theorem, \begin{flalign*}
G_{\y}(\Omega)&=\int_{\Theta}\left(\int_{\Omega} \frac{p(Y,\mathbf{y},\theta)}{p(\mathbf{y},\theta)}dY\right){d\Pi[\theta|\eta(\y)]}=\int_{\Theta} \frac{p(\mathbf{y},\theta)}{p(\mathbf{y},\theta)}d\Pi[\theta|\eta(\y)]=\Pi[\Theta|\eta(\y)]=1
\end{flalign*}

\end{proof}

\subsection{Theorem \ref{thm2}}


\begin{proof}
Define $\rho_{H}$ to be the Hellinger metric, {that is,} for absolutely
continuous probability measures $P$ and $G$,
\[
\rho_{H}\{P,G\}=\left\{  \frac{1}{2}\int\left[  \sqrt{{dP}}-\sqrt{{dG}%
}\right]  ^{2}d\mu\right\}  ^{1/2},\quad0\leq\rho_{H}\{P,G\}\leq1
\]
for $\mu$ the Lebesgue measure, and define $\rho_{TV}$ to be the total
variation metric,
\[
\rho_{TV}\{P_{{}},G_{{}}\}=\sup_{B\in\mathcal{F}}|P(B)-G(B)|,\quad0\leq
\rho_{TV}\{P,G\}\leq2
\]
Recall that, according to the definition of merging in Blackwell and Dubins
(1962), two predictive measures $P_{\mathbf{y}}$ and $G_{\mathbf{y}}$ are said
to merge if
\[
\rho_{TV}\{P_{\mathbf{y}},G_{\mathbf{y}}\}=o_{\mathbb{P}}(1).
\]

Fix $\epsilon>0$ and define the set $V_{\epsilon}:=\{\mathbf{\theta}\in
\Theta:\rho_{H}\{F_{\mathbf{y}},P(\cdot|{\mathbf{y}},\mathbf{\theta
})\}>\epsilon/2\}$. By convexity of $\rho_{H}\{F_{\mathbf{y}},\cdot\}$ and
Jensen's inequality, \begin{flalign*}
\rho_{H}\{F_{\y},P_{\y}\}&\leq \int_{\Theta}\rho_{H}\{F_{\y},P(\cdot|\y,\T)\}d\Pi[\T|\y]\\&\leq \int_{V_{\epsilon}}\rho_{H}\{F_{\y},P(\cdot|\y,\T)\}d\Pi[\T|\y]+\int_{V_{\epsilon}^{c}}\rho_{H}\{F_{\y},P(\cdot|\y,\T)\}d\Pi[\T|\y]\\&\leq \Pi[V_{\epsilon}|\y]+\frac{\epsilon}{2}\Pi[V_{\epsilon}^{c}|\y].
\end{flalign*}By definition, $\mathbf{\theta}_{0}\notin V_{\epsilon}$ and
therefore, by Assumption \ref{ass1} {Part }(4), $\Pi\lbrack V_{\epsilon
}|\mathbf{y}]=o_{\mathbb{P}}(1)$. Hence, we can conclude:
\begin{equation}
\rho_{H}\{F_{\mathbf{y}},P_{\mathbf{y}}\}\leq o_{\mathbb{P}}(1)+\frac
{\epsilon}{2}. \label{hell1}%
\end{equation}

Now, apply the triangle inequality to obtain $\rho_{H}\{P_{\mathbf{y}}^{{}%
},G_{\mathbf{y}}\}\leq\rho_{H}\{F_{\mathbf{y}},P_{\mathbf{y}}\}+\rho
_{H}\{F_{\mathbf{y}},G_{\mathbf{y}}^{{}}\}.$ Using \eqref{hell1}, convexity of
$\rho_{H}$, and Jensen's inequality, \begin{flalign*}
\rho_{H}\{P_{\y},G_{\y}\}&\leq o_{\mathbb{P}}(1)+\frac{\epsilon}{2} +\int_{\Theta}\rho_{H}\{{F}_{\y}, P(\cdot|\y,\T)\}d\Pi[\T|\eta(\y)]\\&\leq o_{\mathbb{P}}(1)+\frac{\epsilon}{2}+\int_{V_{\epsilon}^{c}}\rho_{H}\{{F}_{\y}, P(\cdot|\y,\T)\}d\Pi[\T|\eta(\y)]+\int_{V_{\epsilon}^{}}\rho_{H}\{{F}_{\y}, P(\cdot|\y,\T)\}d\Pi[\T|\eta(\y)]\\&\leq o_{\mathbb{P}}(1)+\frac{\epsilon}{2}+\frac{\epsilon}{2} \Pi[V_{\epsilon}^{c}|\eta(\y)]+\Pi[V_{\epsilon}|\eta(\y)].
\end{flalign*}From the Bayesian consistency of $\Pi\lbrack\cdot|\eta
(\mathbf{y})]$, for any $\epsilon^{\prime}\leq\epsilon$, $A_{\epsilon^{\prime
}}^{c}\not \subset \limsup_{T\rightarrow\infty}V_{\epsilon}$,\textbf{ }where
we recall that the set $A_{\epsilon}$ was defined previously as\textbf{
}$A_{\epsilon}:=\{\theta\in\Theta:\Vert\mathbf{\theta}-\mathbf{\theta}%
_{0}\Vert>\epsilon\}$. Applying again Assumption \ref{ass1} {Part }(4),
$\Pi\lbrack V_{\epsilon}|\eta(\mathbf{y})]=o_{\mathbb{P}}(1)$, and we can
conclude \begin{flalign*}
\rho_{H}\{P_{\y}^{},G^{}_{\y}\}&\leq o_{\mathbb{P}}(1)+\epsilon.
\end{flalign*}

For probability distributions $P,G$, recall that \begin{flalign*}
0\leq\rho_{TV}\{P,G\}\leq\sqrt{2}\cdot\rho_{H}\{P,G\}.
\end{flalign*}Applying the relationship between $\rho_{H}$ and $\rho_{TV}$,
yields the stated result.
\end{proof}

\subsection{Theorem \ref{thm1}}

\begin{proof}

\noindent\textbf{\underline{Part (i)}:} Under correct model specification, for
any $B\in\mathcal{F}_{T+1}$ \begin{flalign}
P_{\y}(B)&=\int_{\Omega}\int_{\Theta}^{}p(Y|\y,\theta)d\Pi[\T|\y]d\delta_{Y}(B)\nonumber\\&=\int_{\Omega}\int_{\Theta}p(Y|\y,\theta)d\delta_{\T}(\T_0)d\delta_{Y}(B)+\int_{\Omega}\int_{\Theta}p(Y|\y,\T)\{d\Pi[\T|\y]-d\delta_{\T}(\T_0)\}d\delta_{Y}(B)\nonumber\\&=F_{\y}(B)+\int_{\Omega}\int_{\Theta}p(Y|\y,\T)d\delta_{Y}(B)\{d\Pi[\T|\y]-d\delta_{\T}(\T_0)\}.\label{decomp}
\end{flalign}The second term in equation \eqref{decomp}, $\int_{\Theta}%
\int_{\Omega}p(Y|\mathbf{y},\mathbf{\theta})d\delta_{Y}(B)\{d\Pi
\lbrack\mathbf{\theta}|\mathbf{y}]-d\delta_{\mathbf{\theta}}(\mathbf{\theta
}_{0})\}$, is a bounded and continuous function of $\mathbf{\theta}$ for each
$\mathbf{y}$. Therefore, from the posterior concentration of $\Pi
[\mathbf{\theta}|\mathbf{y}]$ to $\delta_{\theta_{0}}$, Assumption \ref{ass1}
part (4),
\[
\int_{\Omega}\int_{\Theta}p(Y|\mathbf{y},\mathbf{\theta})d\delta_{Y}%
(B)\{d\Pi[\mathbf{\theta}|\mathbf{y}]-d\delta_{\mathbf{\theta}}(\mathbf{\theta
}_{0})\}=o_{\mathbb{P}}(1)
\]
and it follows that $P_{\mathbf{y}}=F_{\mathbf{y}}+o_{\mathbb{P}}(1)$.
Applying this result to $\mathbb{M}(P_{\mathbf{y}},F_{\mathbf{y}})$ we can
conclude \begin{flalign*}
\mathbb{M}(P_{\y},F_{\y})=\int_{\Omega}S(P_{\y},Y)dF_{\y}(Y)&=\int_{\Omega}S(F_{\y},Y)dF_{\y}(Y)+o_{\mathbb{P}}(1).
\end{flalign*}

The same derivations to the above yield that, under Assumption \ref{ass1} part
(4), $G_{\mathbf{y}}=F_{\mathbf{y}}+o_{\mathbb{P}}(1)$, and \begin{flalign*}
\mathbb{M}(G_{\y},F_{\y})=\int_{\Omega}S(G_{\y},Y)dF_{\y}(Y)&=\int_{\Omega}S(F_{\y},Y)dF_{\y}(Y)+o_{\mathbb{P}}(1).
\end{flalign*}Therefore,
\[
\mathbb{M}(P_{\mathbf{y}},F_{\mathbf{y}})-\mathbb{M}(G_{\mathbf{y}%
},F_{\mathbf{y}})=o_{\mathbb{P}}(1).
\]


\noindent\textbf{{\underline{Part (ii)}:}} Define the random variables,
$\hat{Y}=S(P_{\mathbf{y}},Y_{T+1})$ and $\hat{X}=S(G_{\mathbf{y}},Y_{T+1})$.
The result of\textbf{ Part (i) }can then be stated as, up to an $o_{\mathbb{P}%
}(1)$ term, $\mathbb{E}\left[  \hat{Y}{|}\mathbf{y}\right]  =\mathbb{E}\left[
\hat{X}{|}\mathbf{y}\right]  $. Therefore, up to an $o(1)$ term,%
\[
\mathbb{E}\left[  \hat{Y}\right]  =\mathbb{E}\left[  \mathbb{E}\left[  \hat
{Y}|\mathbf{y}\right]  \right]  =\mathbb{E}\left[  \mathbb{E}\left[  \hat
{X}|\mathbf{y}\right]  \right]  =\mathbb{E}\left[  \hat{X}\right]  .
\]

\bigskip

\noindent\textbf{{\underline{Part (iii)}:}}
For $\eta_{0}=\eta(\mathbf{y})$, rewrite $g(y_{T+1}|\mathbf{y})$ as
\begin{flalign*}
g(y_{T+1}|\mathbf{y})&=\int_{\Theta}\frac{p(y_{T+1},\theta,\mathbf{y})}{p(\theta,\mathbf{y})}\frac{p(\eta_0|\theta)p(\theta)}{\int_{\theta}p(\eta_0|\theta)p(\theta)d\theta}d\theta= \int_{\Theta}\frac{p(y_{T+1},\theta,\mathbf{y})}{p(\mathbf{y}|\theta)p(\theta)}\frac{p(\eta_0|\theta)p(\theta)}{\int_{\theta}p(\eta_0|\theta)p(\theta)d\theta}d\theta\\&=\int_{\Theta}\frac{p(y_{T+1},\theta,\mathbf{y})}{\int_{\theta}p(\eta_0|\theta)p(\theta)d\theta}\frac{p(\eta_0|\theta)}{p(\mathbf{y}|\theta)}d\theta
\end{flalign*}Likewise, $p(y_{T+1}|\mathbf{y})$ can be rewritten as
$p(y_{T+1}|\mathbf{y})=\int_{\Theta}{p(y_{T+1},\mathbf{\theta},\mathbf{y}%
)}d\mathbf{\theta}/{\int_{\theta}p(\mathbf{y}|\mathbf{\theta})p(\mathbf{\theta
})d}\mathbf{\theta}.$ The result follows if and only if $p(\mathbf{y}%
|\mathbf{\theta})=p(\mathbf{y})p(\eta_{0}|\mathbf{\theta})$.
\end{proof}

\subsection{Posterior Consistency in the INAR(1) and MA(2) Examples}

Under the assumption of correct model specification, posterior consistency in
ABC can be demonstrated by verifying the sufficient conditions given in
Theorem 1 of \cite{FMRR2016}, which we restate here for ease of
exposition:\medskip

\noindent\textbf{[A1]} There exist a continuous, injective map $b:\Theta
\rightarrow\mathcal{B}\subset\mathbb{R}^{k_{\eta}}$ and a function $\rho
_{T}(\cdot)$ satisfying: $\rho_{T}(u)\rightarrow0$ as $T\rightarrow\infty$
for all $u>0$, and $\rho_{T}(u)$ monotone non-increasing in $u$ (for any given
$T$), such that, for all $\theta\in\Theta$,
\[
P_{{\theta}}\left[  d_{}\{{\eta}(\mathbf{z}),{b}({\theta})\}>u\right]  \leq
c({\theta})\rho_{T}(u),\quad\int_{\Theta}c({\theta})d\Pi({\theta})<+\infty
\]
where either of the following is satisfied:

\begin{enumerate}
\item[\textbf{(i)}] \textit{Polynomial deviations:} There exist a positive
sequence $v_{T}\rightarrow+\infty$ and $u_{0},\kappa>0$ such that $\rho
_{T}(u)=v_{T}^{-\kappa}u^{-\kappa}$, for $u\leq u_{0}$.

\item[\textbf{(ii)}] \textit{Exponential deviations:} There exists $h_{\theta
}(\cdot)>0$ such that $P_{\theta}[d\{\eta(\mathbf{z}),b(\theta)\}>u]\leq
c(\theta)e^{-h_{\theta}(u v_{T})}$ and there exists $c,C>0$ such that
\[
\int_{\Theta} c(\theta)e^{-h_{\theta}(u v_{T})}d\Pi(\theta)\leq Ce^{-c(u
v_{T})^{\tau}},\;\;\text{for }u\leq u_{0}.
\]

\end{enumerate}


\noindent\textbf{[A2]} The prior $p(\theta)$ is absolutely continuous with
respect to the Lebesgue measure and satisfies $p(\theta^{0}) > 0$.\medskip


\subsubsection{INAR(1)}

Recall the INAR(1) model
\[
y_{t}=\sum_{j=0}^{y_{t-1}}B_{j}(\rho)+\varepsilon_{t},\label{new_inar}%
\]
where $B_{j}(\rho)$ are $i$.$i$.$d$ are Bernoulli random variables with
probability $\rho$, and $\varepsilon_{t}$ is $i$.$i$.$d$ Poisson with
intensity parameter $\lambda$. The summary statistics chosen for this example
were the sample mean, $\bar{y}$, and the first three sample autocovariances.

The parameters are $\theta=(\rho,\lambda)^{\prime}$ and our prior space is
uniform over
\[
\Theta:=\{\theta\in\Theta:\rho\in\lbrack0,1-\delta],\;\lambda\in
\lbrack0,10]\},
\]
for some small $\delta>0$. The uniform prior $p(\theta)$ over $\Theta$
automatically fulfills Assumption \textbf{[A2]} for $\theta_{0}$ in this space.

For any $\theta\in\Theta$ and $\mathbf{z}$ simulated from \eqref{new_inar},
it follows that
\[
\eta(\mathbf{z})=%
\begin{pmatrix}
\bar{z}\\
\gamma_{1}\\
\gamma_{2}\\
\gamma_{3}%
\end{pmatrix}
=%
\begin{pmatrix}
\lambda/(1-\rho)\\
\rho\\
\rho^{2}\\
\rho^{3}%
\end{pmatrix}
+o_{P}(1).
\]
Define $b(\theta)=(\lambda/(1-\rho),\rho,\rho^{2},\rho^{3})^{\prime}$ and note
that $b(\theta)$ is continuous. The linear allocation of $\rho$ as the second
element of $b(\theta)$ ensures that $\theta\mapsto b(\theta)$ is injective in
$\theta$. From the structure of $\eta(\mathbf{z})$, $V(\theta)=\mathbb{E}%
[\{\eta(\mathbf{z})-b(\theta)\}\{\eta(\mathbf{z})-b(\theta)\}^{\prime}]$
satisfies $\text{tr}\{V(\theta)\}<\infty$ for all $\theta\in\Theta$. By
Markov's inequality, 
\begin{flalign*}
P_{\theta}\left\{\|\eta(\z)-b(\theta)\|>u\right\}=P_{\theta}\left\{\|\eta(\z)-b(\theta)\|^{2}>u^2\right\}&\leq \frac{\text{tr}\{V(\theta)\}}{u^2 T}.
\end{flalign*}As a result, Assumption \textbf{[A1]} is satisfied with
$\rho_{T}(u)=1/(T^{1/2}u)^{2}$.

\subsubsection{MA(2)}

We now verify the conditions in the moving average model of order two:%
\[
y_{t}=e_{t}+\theta_{1}e_{t-1}+\theta_{2}e_{t-2}\;\;(t=1,\dots,T),
\]
where $\{e_{t}\}_{t=1}^{T}$ is a sequence of white noise random variables with
variance $\sigma^{2}$ such that, for some $\delta>0$, $\mathbb{E}%
[e_{t}^{4+\delta}]<\infty$. Our prior for $\theta=(\sigma^{2},\theta
_{1},\theta_{2})^{\prime}$ is uniform over the following region,
\[
\Theta:=\left\{  \theta\in\Theta:0\leq\sigma^{2}\leq3,\;0\leq\theta_{1}%
\leq1-\delta,\;0\leq\theta_{2}\leq1-\delta\right\}  ,
\]
for some small $\delta>0$. The summary statistics for this exercise are given
by the sample autocovariances $\eta_{j}(\mathbf{y})=T^{-1}\sum_{t=1+j}%
^{T}y_{t}y_{t-j}$, for $j=0,1,\dots,l$.
For any $\theta\in\Theta$, $\eta_{j}(\mathbf{z})=T^{-1}\sum_{t=1+j}^{T}%
z_{t}z_{t-j}$. Define $b_{j}(\theta)=\mathbb{E}_{\theta}(z_{t}z_{t-j})$ and
take $b(\theta)=(b_{0}(\theta),b_{1}(\theta),b_{2}(\theta))^{\prime}$. Each
choice of the summary statistics in the MA(2) example that leads to a Bayesian
consistent posterior has these components in common.

Firstly, note that $\theta\mapsto b(\theta)=(\sigma^{2}(1+\theta_{1}%
^{2}+\theta_{2}^{2}),(1+\theta_{2})\theta_{1},\theta_{2})^{\prime}$ is
continuous in $\theta$. In addition, from the linear allocation of $\theta_2$ in $b_2(\theta)$, it follows that $\theta\mapsto
b(\theta)$ is injective over $\Theta$. Now, take $d_{2}\{\eta(\mathbf{z}),b(\theta
)\}=\left\Vert \eta(\mathbf{z})-b(\theta)\right\Vert $ for simplicity. Under
the moment restriction on $e_{t}$ above, $V(\theta)=\mathbb{E}[\{\eta
(\z)-b(\theta)\}\{\eta(\z)-b(\theta)\}^{\prime}]$ satisfies $\text{tr}%
\{V(\theta)\}<\infty$ for all $\theta\in\Theta$. By Markov's inequality,\begin{flalign*}
P_{\theta}\left\{\|\eta(\z)-b(\theta)\|>u\right\}=P_{\theta}\left\{\|\eta(\z)-b(\theta)\|^{2}>u^2\right\}&\leq \frac{\text{tr}\{V(\theta)\}}{u^2 T}+o(1/T),
\end{flalign*}where the $o(1/T)$ term comes from the fact that there are
finitely many non-zero covariance terms due to the $m$-dependent nature of the
series. As a result, Assumption \textbf{[A1]} is satisfied with $\rho
_{T}(u)=1/(T^{1/2}u)^{2}$.

The uniform prior $p(\theta)$ automatically fulfills Assumption \textbf{[A2]}
for $\theta_{0}$ in this space.

\end{document}